\documentclass[reqno]{amsart}
\usepackage[utf8]{inputenc}
\usepackage{pgffor}
\usepackage{xcolor, soul}
\definecolor{hyperlink}{rgb}{0.7 0 0}
\usepackage[colorlinks=true,allcolors=hyperlink]{hyperref}
\usepackage{booktabs}
\usepackage{mathtools, mathdots, amsthm, amssymb, mathrsfs}
\usepackage{tikz}
\usetikzlibrary{decorations.pathmorphing, patterns,shapes}
\usepackage{graphicx}
\usepackage{cases}
\usepackage{array}
\usepackage{enumitem}
\usepackage[figureposition=top, skip=3pt]{caption}
\usepackage{caption, subcaption}
\usepackage{natbib}

\graphicspath{ {./images/} }

\usepackage[vlined,boxed,commentsnumbered]{algorithm2e} 

\usepackage{listings}
\usepackage[scaled]{beramono}
\usepackage[T1]{fontenc}

\newcommand{\zigzag}[1]{\begin{tikzpicture}
  \draw[decoration = {zigzag, segment length = 3mm, amplitude = 0.5mm},decorate, color = codegrey] (0,0.1)--(#1,0.1);
\end{tikzpicture}
}

\definecolor{codegreen}{rgb}{0,0.6,0}
\definecolor{codegrey}{rgb}{0.5,0.5,0.5}
\definecolor{codepurple}{rgb}{0.58,0,0.82}
\definecolor{backcolor}{rgb}{0.95,0.95,0.92}

\usepackage{mdframed}
\newmdenv[
  topline=false,
  bottomline=false,
  rightline=false,
  skipabove=\topsep,
  skipbelow=\topsep
]{leftrule}

\newcommand{\rd}{\mathrm d}

\newcommand{\eqd}{\stackrel{\mathrm{dist.}}{=}}

\newcommand{\cond}{\xrightarrow{\mathrm{dist.}}}

\newcommand{\tos}{\stackrel{S}{\rightarrow}}
\newcommand{\conp}{\xrightarrow{\mathrm{prob.}}}

\newcolumntype{L}{>{$}l<{$}}

\makeatletter

\DeclareCaptionSubType*{figure}
\DeclareCaptionSubType*{table}
\newtheorem{lemma}{Lemma}
\newtheorem{proposition}{Proposition}
\newtheorem{theorem}{Theorem}

\theoremstyle{definition}

\newtheorem*{assumption}{Assumption}

\theoremstyle{remark}
\newtheorem{remark}{Remark}[theorem]
\newtheorem*{remark*}{Remark}

\newtheoremstyle{named}{}{}{\itshape}{}{\bfseries}{.}{.5em}{\thmnote{#3 }#1}
\theoremstyle{named}
\newtheorem*{namedtheorem}{Theorem}

\renewenvironment{proof}[1][\proofname]
    {\begin{leftrule}   
        \par        \normalfont \topsep6\p@\@plus6\p@        \trivlist
                        \item[\hskip\labelsep\bf\itshape
              #1.]\ignorespaces
    }
    {\qed\endtrivlist\end{leftrule}}
    
\renewcommand{\vec}{\mathbf}

\newcommand{\customlabel}[2]{#2\def\@currentlabel{#2}\label{#1}}

\DeclarePairedDelimiter{\floor}{\lfloor}{\rfloor}

\DeclareMathOperator{\E}{\mathbb E}

\undef\P
\DeclareMathOperator{\P}{\mathbb P}
\DeclareMathOperator{\Q}{\mathbb Q}
\DeclareMathOperator{\I}{\mathbb I}

\newcommand{\R}{\mathbb R}

\tikzstyle{level 1}=[level distance=2.5cm, sibling distance=3cm,->]
\tikzstyle{level 2}=[level distance=2.5cm, sibling distance=1.1cm,->]
\tikzstyle{level 3}=[level distance=2.5cm, sibling distance=1.1cm,->]

\tikzstyle{bag} = [text width=2em, text centered]
\tikzstyle{end} = []

\makeatother

\calclayout

\begin{document}
\title
    {Risk-Neutral Pricing of Random-Expiry Options Using Trinomial Trees}
\author[S\'ebastien Bossu \and Michael Grabchak]{S\'ebastien Bossu \and Michael Grabchak{*}\\ {\em Department of Mathematics \& Statistics, UNC Charlotte}}
\thanks{\noindent* 
Email: mgrabcha@charlotte.edu.\\  
\indent Our Python code is available online at \href{https://github.com/sbossu/Bossu-Grabchak_Trinomial}{github.com/sbossu/Bossu-Grabchak\_Trinomial}.  
}

\date{\today}

\begin{abstract}
    Random-expiry options are nontraditional derivative contracts that may expire early based on a random event.  We develop a methodology for pricing these options using a trinomial tree, where the middle path is interpreted as early expiry.  We establish that this approach is free of arbitrage, derive its continuous-time limit, and show how it may be implemented numerically in an efficient manner.
\end{abstract}

\maketitle



\section{Introduction}

We introduce and develop methods for the pricing of \emph{random-expiry} (RE) options\footnote{As is commonly done in the literature, we use the terms \emph{derivative, option, payoff}, and \emph{contingent claim} interchangeably.}, which are options that expire at some random time $\tau$. For instance, $\tau$ may represent the time of an insurable event such as death, surgery, or unemployment. In other situations, it may be the time of, e.g., a sports or political victory, a peace agreement, or a company merger. A simple example of an RE option is a contingent claim that delivers one share of stock at time $\tau$ or at the maturity date, whichever comes sooner. Since a derivative contract whose payoff matches the underlying asset is known to practitioners as a zero-strike call, we refer to the aforementioned claim as a \emph{random-expiry zero-strike call} (RE-ZSC) option.  More generally, an RE option has an arbitrary payoff function $f(S)$ paid at time $\tau$ but no later than the maturity date, where $S$ is the price of the underlying asset at expiration.

In practice, RE options encompass a wide range of situations, including: variable life insurance products; insurance-linked securities such as `CAT bonds' \citep{AAA:2022}; employee stock options with accelerated vesting triggered by a company merger \citep{mcintosh-harmetz:2002}; benefit payments from stable value funds; bank loans that can be terminated based on balance sheet criteria; callable, puttable, extendible, and convertible instruments \citep{longstaff:1990}; credit derivatives, etc.  We do not discuss the specific details of these potentially intricate instruments in this paper. However, in principle, they should fit into our model with ad hoc extensions and modifications, which will be the basis for future research.

\begin{figure}[hbtp]
    \centering
\begin{tikzpicture}[grow=right, sloped, level 1/.style={level distance=2.5cm, sibling distance=1.5cm,->}]
\usetikzlibrary {arrows.meta} 
\node[bag] {$ S_0  $}
    child {
        node[bag] {$ S_0 d $}        
            child {
                node[end, label=right:
                    {$ S_0 dd $}] {}
                edge from parent
                node[above] {}
                node[below]  {$q_d^{(1)}$}
            }
            child[-{Rays[length=2mm]}] {
                node[end,right=-0.8cm, label=right:
                    {$ S_0 d $}] {}
                edge from parent
                node[above=-0.1cm] {$q_m^{(1)}$}
                node[below]  {}
            }
            child {
                node[end, label=right:
                    {$ S_0 du $}] {}
                edge from parent
                node[above] {$q_u^{(1)}$}
                node[below]  {}
            }
            edge from parent
            node[above] {}
            node[below]  {$q_d^{(0)}$}
    }
child[-{Rays[length=2mm]}] {
        node[bag,right=-0.8cm] {$ S_0  $}        
            edge from parent
            node[above=-0.1cm] {$q_m^{(0)}$}
            node[below]  {}
    }
    child {
        node[bag] {$ S_0 u $}        
        child {
                node[end, label=right:
                    {$ S_0 ud $}] {}
                edge from parent
                node[above] {}
                node[below]  {$q_d^{(1)}$}
            }
            child[-{Rays[length=2mm]}] {
                node[end,right=-0.8cm, label=right:
                    {$ S_0 u $}] {}
                edge from parent
                node[above=-0.1cm] {$q_m^{(1)}$}
                node[below]  {}
            }
            child {
                node[end, label=right:
                    {$ S_0 uu $}] {}
                edge from parent
                node[above] {$q_u^{(1)}$}
                node[below]  {}
            }
        edge from parent        
            node[above] {$q_u^{(0)}$}
            node[below]  {}
    };
\end{tikzpicture}
    \caption{Stock price evolution over a two-step trinomial tree with early expiry on the middle nodes (no dividends). }
    \label{fig:trinomial-early}
\end{figure}
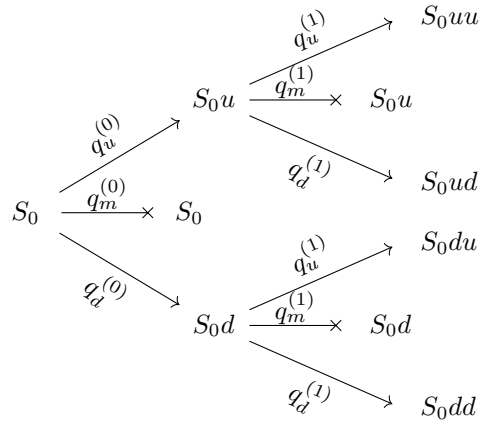

Due to the uncertainty in expiration time, the problem of pricing RE options cannot be addressed with classical methods.  In particular, it is \emph{a priori} unknown whether a risk-neutral pricing measure even exists.  We approach this problem from the ground up and formulate it in an incomplete markets setting. Specifically, we use a trinomial tree structure, where the middle path represents early expiry; see Figure \ref{fig:trinomial-early} for an illustration of the stock price evolution over a two-step trinomial tree with early expiry as prescribed in the RE-ZSC contract.

Our approach relies on embedding this tree in a classical trinomial tree, where middle nodes and their descendants correspond to early payoff values with carried interest to maturity (see Figure \ref{fig:stocks}).  We establish that this approach is free of arbitrage.

We also derive the continuous-time limit of the tree.  We note that the limiting distribution is not Brownian motion as there is an additional random term, which can lead to a variety of distributions. Thus, aside from intrinsic interest in the problem considered here, this paper also contributes to the literature on tree approximations to distributions that are not the normal (Gaussian) distribution. While normal distributions are the foundation on which the classic theory of Black-Scholes option pricing is built, it has long been recognized that it is not a realistic model for financial returns. Instead, it has been suggested that other infinitely divisible distributions give rise to more realistic models; see, e.g.,\ the well-regarded monograph \cite{Cont:Tankov:2004} or the discussion in \cite{Grabchak:Samorodnitsky:2010}. An important limitation in working with such distributions is that the useful tool of binomial tree approximations is not readily available. In fact, \cite{Rachev:Ruschendorf:1995} showed that binomial trees with time-homogeneous parameters can only lead to normal or Poisson limiting distributions. In order to get more interesting distributions in the limit, one needs an additional source of randomness.  In our approach, this additional source is the random expiration time $\tau$ and, by considering a variant of the trinomial tree model, we are able to perform risk-neutral pricing.

In related literature, there have been several attempts at introducing this additional randomness, see, e.g.,\ \cite{Rachev:Samorodnitsky:1993}, \cite{Rachev:Ruschendorf:1995}, \cite{Karandikar:Rachev:1995}, \cite{Rejman:Weron:Weron:1997}, \cite{Ishwaran:Jahandideh:Zarepour:2008}, and the monograph \cite{Rachev:Mittnik:2000}. However, in all of these approaches, the authors are not able to get a fully risk-neutral measure. They get a conditionally risk-neutral measure, given the additional source of randomness, but then they still need to take an expectation over a measure that is not risk-neutral. In a different direction \cite{Kellezi:Webber:2004} and \cite{Maller:Solomon:Szimayer:2006} proposed multinomial tree approximations, where the degree of the tree approaches infinity. 

In addition to the literature cited above, we provide a selection of sources on {binomial and} trinomial trees and incomplete markets. Binomial and trinomial trees for option pricing were introduced, respectively, by \cite{Cox-Ross-Rubinstein:1979} and \cite{Boyle:1986}. Pricing using more general multinomial trees is studied in \cite{Madan:Milne:Shefrin:1989}. \citet{Duffie:1996} provides an excellent summary of the results and challenges encountered by researchers working on incomplete markets with a countably infinite or a continuous state space.  \cite{Wolczynska:1998} investigates risk-minimization techniques for trinomial option pricing in incomplete markets. \cite{Jacod-Protter:2017} consider incomplete market models with an infinite number of risk neutral measures. \citet{kim-stoyanov-rachev-fabozzi:2019} introduce a risk-neutral trinomial tree and derive a hedging strategy based on an additional perpetual derivative used as a second asset for hedging. \citet{Jarrow:2023} and \citet{grigorian-jarrow:2024} propose a methodology to uniquely price non-traded assets using no-arbitrage in an incomplete market, which can be applied to insurance contracts and other illiquid assets.  

The rest of this paper is organized as follows. Section~\ref{sec:theoretical-background} reviews the elements of asset pricing theory that provide the foundation for our approach.  Section~\ref{sec:trinomial} introduces the trinomial tree that we use to model RE options, and Section~\ref{sec:RE-pricing} gives our main theoretical results for pricing. Section~\ref{sec:continuous} derives the continuous-time limit of the trinomial tree and makes a connection with subordinated Brownian motion and related infinitely divisible distributions. Section~\ref{sec:numerical} introduces and compares three algorithms to price RE options and discusses numerical results. Finally, Section~\ref{sec:conclusions} presents our conclusions and directions for further research. For ease of reading, all proofs are deferred to the appendix, where one can also find the Python code implementation of our pricing algorithms. More detailed Python code can be found on GitHub.

Before proceeding, we establish our standing notation. We say that a vector $\vec x$ is strictly positive and write $\vec x > 0$ if all of its coefficients are strictly positive. For any two real numbers $x,y$, we write $\max(x,y)$ for their maximum and $x \wedge y$ or $\min(x,y)$ for their minimum.  For any function $f$ we write $(\min_x, \max_x) f(x)$ for the open interval from the minimum value to the maximum value of $f$. We write $\E^{\Q}$ for the expectation operator under probability measure $\Q$. For any event $A$, we write $\I_A$ to denote the indicator function on $A$. We use $\circ$ to denote the composition of functions operator, we write $:=$ to denote a defining equality, and we denote the floor function by $\floor{\cdot}$. Any void sum is deemed to vanish. 

\section{Theoretical background} \label{sec:theoretical-background}

In this section we review the standard theory upon which our results are built and introduce the notation that we use throughout this paper. While there are many references for these topics, we mostly follow \citet{cerny}.

\subsection{Asset pricing in finite incomplete markets}

A one-period Arrow-Debreu finite market has $\ell$ tradable assets and $n$ states of the world $\Omega := \{ \omega_i \}_{1\leq i\leq n}$. The terminal asset prices are represented by the $n\times \ell$ matrix $\vec A := (a_{i,j})_{1\leq i\leq n,1\leq j\leq \ell}$, where $a_{i,j}$ is the terminal price of asset $j$ in state of the world $\omega_i$. We write $\vec A =: [\vec a_1, \dots, \vec a_\ell]$ and note that column $\vec a_j\in\R^n$ is the payoff of asset $j$. The market is complete when any asset with payoff vector $\vec v := (v_i)_{1\leq i\leq n}$ can be replicated by some portfolio $\vec x := (x_j)_{1\leq j \leq \ell}$ of the tradable assets, i.e.,\ when the equation $\vec {Ax} = \vec v$ always has a solution.  This can only happen when $\ell \geq n$ and $\vec A$ is of full row rank $n$.  In particular, if $\ell=n$, then the matrix is square and the replicating portfolio is uniquely determined by $\vec x = \vec A^{-1}\vec v$.  If, instead, $\ell>n$, then there are $\ell-n$ redundant assets that can be removed to revert to the $\ell=n$ case.
When $\ell<n$, the market is incomplete and there are infinitely many payoffs $\vec v\in\R^n$, which cannot be replicated by any portfolio $\vec x\in\R^\ell$ of the tradable assets.

In this theory, there are two payoffs of particular interest:
\begin{itemize}[leftmargin=*,parsep=6pt]
    \item \emph{Arrow-Debreu securities} $\vec e_i := (0, \dots, 0, 1, 0, \dots, 0)^T$, which pay \$1 in state of the world $\omega_i$ and 0 otherwise. Mathematically, the collection $\{\vec e_i\}_{1\leq i\leq n}$ of Arrow-Debreu securities forms the canonical basis of $\R^n$.  In a complete market, Arrow-Debreu securities exist as replicable payoffs, and if $\ell=n$ the columns of $\vec A^{-1}$ correspond to their respective replicating portfolios.  In an incomplete market with $\ell < n$, at least $n - \ell$ Arrow-Debreu securities are nonreplicable.
    \item The~\emph{risk-free bond} $\vec e := (1, \dots, 1)^T$, which always pays \$1.  Mathematically $\vec e = \sum_i \vec e_i$, i.e.,\ the risk-free bond is the sum of all Arrow-Debreu securities.  In a complete market the risk-free bond is always replicable and, when  $\ell = n$, the replicating portfolio is $\vec x = \vec A^{-1}\vec e$.  In an incomplete market of purely risky tradable assets, it is typically nonreplicable.
\end{itemize}

So far we have only discussed \emph{terminal} asset prices but not \emph{current} prices, probabilities, or arbitrage.  The latter concepts are diachronic in nature, whereas market completeness is synchronic. We now introduce the vector $\vec s := (s^{(j)})_{1\leq j\leq \ell}$ of \emph{current} spot prices of the tradable assets. With this, we also introduce the possibility of arbitrage if an asset is redundant and mispriced, or, more generally, if a nonnegative payoff can be replicated by a portfolio with a negative price.  Famously \citep[see e.g.][p.42]{cerny}:

\begin{namedtheorem}[Arbitrage]
There is no arbitrage if and only there exists a strictly positive vector $\boldsymbol{\pi} := (\pi_i)_{1\leq i\leq n}$ that is consistent with the tradable asset spot prices, i.e.,\ $\vec s = \vec A^T \boldsymbol{\pi}$.
\end{namedtheorem}

We call $\boldsymbol{\pi}$ a \emph{state-price vector} because $\pi_i > 0$ is the price of the Arrow-Debreu security $\vec e_i$. Given a state-price vector and any payoff $\vec v$, a no-arbitrage price is given by
\[
    v := \boldsymbol\pi^T\vec v = \sum_i \pi_i v_i. 
\]
In particular, a no-arbitrage price for the risk-free bond is given by $b := \boldsymbol\pi^T\vec e = \sum_i \pi_i > 0 $, i.e.,\ it is the sum of all of the state prices. The normalized state-price vector $b^{-1} \boldsymbol{\pi}=:\vec q=(q_i)_{1\leq i\leq n}$ defines a discrete probability measure by  $\Q(\{\omega_i\}) = q_i = \pi_i / b$, $\omega_i\in\Omega$.  In addition, any asset with payoff vector $\vec v$ defines a random variable $V$ by $V(\omega_i) = v_i$, $\omega_i\in\Omega$, which corresponds to the future asset price in each state of the world. The price of this arbitrary asset is given by $v = b \E^{\Q}(V)= b \sum_i q_i v_i $, which is the expected payoff \emph{with risk-free discounting}.  The measure $\Q$ is called \emph{risk-neutral} because the expected risk premium is $\E^{\Q}(\frac{V-v}v) - \frac{1-b}b = 0$; equivalently the expected return on any risky asset is equal to the risk-free rate.

With this concept in hand, there is no arbitrage if and only if a risk-neutral probability measure $\Q$ exists. In an arbitrage-free complete market, the state-price vector $\boldsymbol{\pi}$ and thus the risk-neutral measure $\Q$ are unique. In an arbitrage-free incomplete market, there are infinitely many state-price vectors $\boldsymbol{\pi}$ and associated risk-neutral measures $\Q$. To obtain these, we must solve: 
\begin{equation}        \label{eq:stateprice}
    \text{Find }\boldsymbol{\pi}>0 \text{ such that } \vec A^T \boldsymbol{\pi} = \vec s.
\end{equation}
Equivalently, in the language of probability:
\begin{equation}        \label{eq:RN}
    \text{Find }b>0 \text{ and } q_1,q_2,\dots, q_n\in(0,1)
    \text{ with } \Sigma_i q_i=1
    \text{ such that } 
    b \E^{\Q}(S^{(j)}) = s^{(j)}, j= 1, 2, \dots, \ell,
\end{equation}
where $S^{(j)}$ is the random variable corresponding to the payoff vector $\vec a_j$, i.e., $S^{(j)}(\omega_i) = a_{i,j}$, and $\Q$ is a probability measure on $\Omega$ with $\Q(\{\omega_i\}) = q_i$. Any state price vector satisfying \eqref{eq:stateprice} (or, equivalently, any risk-neutral measure satisfying \eqref{eq:RN}) can be used for pricing. Thus, any nonreplicable payoff $\vec v$ admits a range of no-arbitrage prices given by
\[
    v\in\left\{ \boldsymbol{\pi}^T \vec v : \boldsymbol{\pi} > 0, \vec A^T \boldsymbol{\pi} = \vec s \right\}.
\]
For arbitrage-free pricing purposes, we may select any such state-price vector $\boldsymbol{\pi}$ or, equivalently, any such risk-neutral measure $\Q$. 
A variety of methodologies for selecting these can be found in the literature, see, e.g., \citet[pp.\ 29--36]{cerny}, \citet{bouzianis-hughston:2020}, and the references therein.

\subsection{Trinomial model}        \label{sec:trinomial-model}

Consider an incomplete market model with $n=3$ states of the world (up, middle, and down) and $\ell=2$ tradable assets: a risk-free bond B with continuous interest rate $r \geq 0$ and a risky stock S with continuous dividend yield $y \geq 0$. The states of the world determine asset prices after a time period $\Delta t$. We denote the bond and stock spot prices as $b = e^{-r\Delta t}$ and $S_0>0$, respectively. Terminal asset payoffs are given by the $3\times2$ matrix
\[
    \vec A = \begin{pmatrix} | & | \\ \vec a_1 & \vec a_2 \\ | & | \end{pmatrix} = \begin{pmatrix} 1 & S_0 u\: e^{y\Delta t} \\ 1 & S_0 m\:e^{y\Delta t} \\ 1 & S_0 d\:e^{y\Delta t} \end{pmatrix}
\]
where $u > m > d > 0$ are ex-dividend stock price movement factors.  In the absence of arbitrage, the state-price and risk-neutral probability vectors are collinear as $\boldsymbol{\pi} = b\vec q$, and they must solve \eqref{eq:RN}, i.e.
\begin{equation}    \label{eq:trinomial-RN}
    \begin{cases}
        q_u + q_m + q_d = 1 \\
        q_u u + q_m m + q_d d = e^{(r-y)\Delta t} \\
        q_m, q_u, q_d > 0.
    \end{cases}
\end{equation}

It is easy to show that the above system has solutions if and only if $ u > e^{(r-y)\Delta t} > d$, in which case they must satisfy
\begin{equation}    \label{eq:RN-trinomial}
    \begin{cases}
        q_u = \frac{e^{(r-y)\Delta t}-d - q_m (m-d)}{u-d} \\
        0 < q_m < \min \left( \frac{e^{(r-y)\Delta t}-d}{m-d}, \frac{u-e^{(r-y)\Delta t}}{u-m} \right)  \\
        q_d =  \frac{u-e^{(r-y)\Delta t} - q_m (u-m)}{u-d}.
    \end{cases}
\end{equation}
Observe that, as $q_m \to 0$, we have $q_u \to \frac{e^{(r-y)\Delta t}-d}{u-d}$ and $q_d \to \frac{u-e^{(r-y)\Delta t}}{u-d}$, and we recover the binomial risk-neutral probabilities \citep[see][p.117]{cerny}.  Observe further that if $ m = e^{(r-y)\Delta t}$, then we may choose any $ 0 < q_m < 1 $.

\section{A trinomial model for random-expiry options}   \label{sec:trinomial}

\subsection{One-period trinomial model}     \label{sec:one-period-trinomial}

Consider a \emph{random-expiry zero-strike call} (RE-ZSC) with the following mechanism: upon signature, it is randomly decided whether the contract delivers one share of stock immediately with probability $p\in(0,1)$ or after one time period of length $\Delta t$ with probability $1-p$. We can describe this situation  using a trinomial model with $n=3$ states of the world, where we interpret the middle path as early expiry. These states and their associated risk-neutral probabilities are given in Table \ref{table: states of world}. Throughout, we assume that $0 < d<e^{(r-y)\Delta t}<u$, which ensures no arbitrage as discussed in Section \ref{sec:trinomial-model}.  It should be noted that, for our purpose, which is the modeling of RE options, we are unconcerned about the behavior of the stock along the middle path. In fact, the ability of the stock to take this path is just a device introduced to allow us to use the middle path as a proxy for early expiry. From this perspective, $m$ is a tuning parameter that we can select in a manner that is most convenient for modeling.

\begin{table}[b]
\caption{States of the world in a one-period trinomial model for random-expiry options.}\label{table: states of world}
\begin{center}
    \begin{tabular}{|c|l|c|}
         \hline
         \textbf{State} & \multicolumn{1}{c|}{\textbf{Description}} & \textbf{Risk-neutral probability} \\
         \hline
         $\omega_1$ & Deferred delivery, stock goes up by an ex-dividend factor $u$ & $ q_u $ \\
         $\omega_2$ & Immediate delivery, stock moves by an ex-dividend factor $m$ & $q_m$ \\
         $\omega_3$ & Deferred delivery, stock goes down by an ex-dividend factor $d$ & $q_d$ \\
         \hline
    \end{tabular}
\end{center}
\end{table}

In our setup, the RE-ZSC payoff vector is $\vec v = \left(\substack{S_0 u \\ S_0 e^{r\Delta t} \\ S_0 d}\right)$, where $S_0 u$ and $S_0 d$ are, respectively, the ex-dividend stock prices in the up and down states $\omega_1$ and $\omega_3$, and $S_0 e^{r\Delta t}$ is the initial stock price with carried interest in the middle state $\omega_2$. If $m=e^{r\Delta t}$, then this payoff is plainly redundant, as it is collinear with the stock payoff $\vec a_2 = \left(\substack{ S_0 u\:e^{y\Delta t} \\ S_0 m\:e^{y\Delta t} \\ S_0 d\:e^{y\Delta t} }\right)$, and its price must be $V_0 = S_0 \:e^{-y\Delta t}$.  But if $m\neq e^{r\Delta t}$ we may use risk-neutral pricing to get the RE-ZSC price as $ V_0^{\text{RE-ZSC}} = q_u S_0 u e^{-r\Delta t} + q_m S_0 + q_d S_0 d e^{-r\Delta t} $ where $q_u,q_m,q_d$ satisfy \eqref{eq:RN-trinomial}.  After simplification, we get
\[
    V_0^{\text{RE-ZSC}}\in\left\{ S_0(e^{-y \Delta t} + q_m(1-m e^{-r\Delta t})) : 0 < q_m < \min\left(\frac{e^{(r-y)\Delta t}-d}{m-d}, \frac{u-e^{(r-y)\Delta t}}{u-m} \right) \right\}.
\]
If $m \leq e^{(r-y)\Delta t}$, this becomes $V_0^{\text{RE-ZSC}} \in \left(S_0 e^{-y\Delta t}, S_0 e^{-y\Delta t}\frac{ue^{y\Delta t}(1-m e^{-r\Delta t})+u-e^{r\Delta t}}{u-m}\right)$, if $e^{(r-y)\Delta t}\leq m < e^{r\Delta t}$ it becomes $V_0^{\text{RE-ZSC}} \in \left(S_0 e^{-y\Delta t}, S_0 e^{-y\Delta t}\frac{e^{r\Delta t}-d -de^{y\Delta t}(1-m e^{-r\Delta t})}{m-d}\right)$, and if $m > e^{r\Delta t}$ it becomes
\\ $V_0^{\text{RE-ZSC}} \in \left(S_0 e^{-y\Delta t}\frac{e^{r\Delta t}-d -de^{y\Delta t}(1-m e^{-r\Delta t})}{m-d}, S_0 e^{-y\Delta t}\right)$.

Our model can be applied \emph{mutatis mutandi} to a general RE option with arbitrary payoff function $f(S)$, where $S$ is the stock price at time 0 if the middle path is taken and it is the ex-dividend price at time $\Delta t$ otherwise.  The payoff vector of the RE option is then $\vec v = \left(\substack{f(S_0 u) \\ f(S_0)e^{r\Delta t} \\ f(S_0 d)}\right)$ and its risk-neutral price is $ V_0 = q_u f(S_0 u) e^{-r\Delta t} + q_m f(S_0) + q_d f(S_0 d) e^{-r\Delta t} $.

At this point, since we are free to select $q_m$, we can take it to be $p$. However, to do so we need $m$ to be such that 
$p = q_m < \min\left(\frac{e^{(r-y)\Delta t}-d}{m-d}, \frac{u-e^{(r-y)\Delta t}}{u-m} \right)$. Since $m$ is a free parameter, this suggests taking $m=e^{(r-y)\Delta t}$ so that this condition is satisfied for any value of $p = q_m \in (0,1)$.

\subsection{Multi-period trinomial tree}

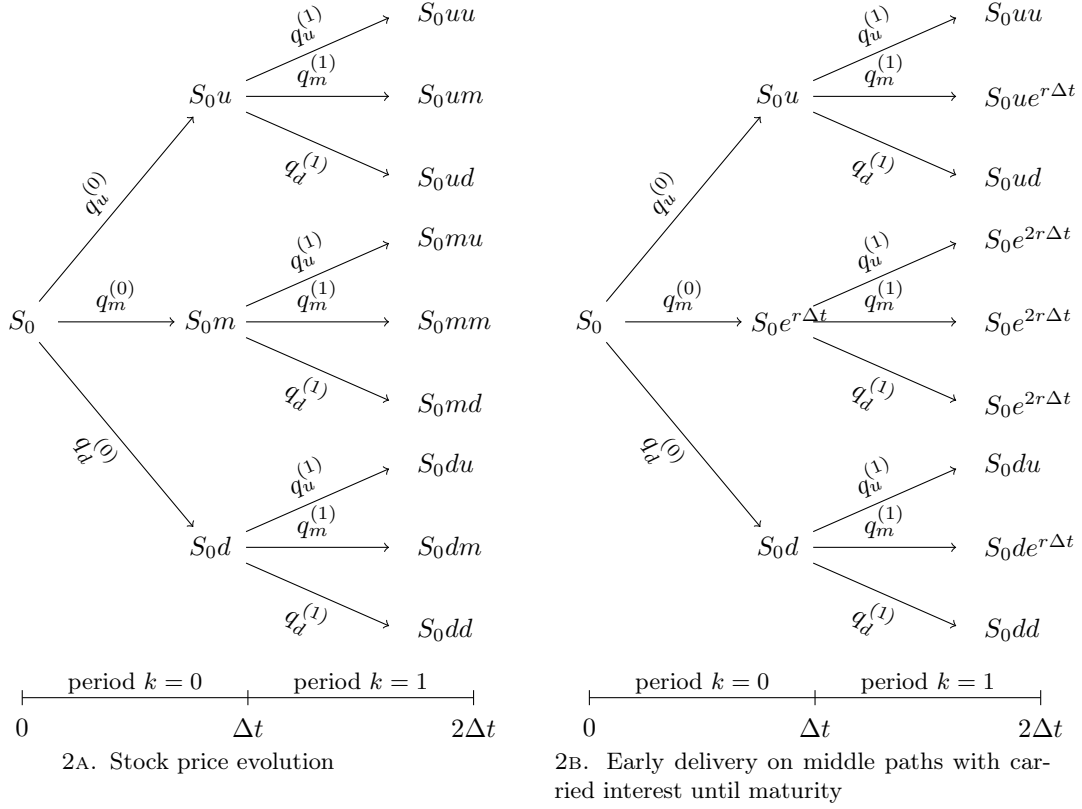
\begin{figure}
     \centering
     \begin{subfigure}[b]{0.4\textwidth}
        \centering
                \begin{tikzpicture}[grow=right, sloped]
        \node[bag] {$ S_0  $}
            child {
                node[bag] {$ S_0 d $}        
                    child {
                        node[end, label=right:
                            {$ S_0 dd $}] {}
                        edge from parent
                        node[above] {}
                        node[below]  {$q_d^{(1)}$}
                    }
                    child {
                        node[end, label=right:
                            {$ S_0 dm $}] {}
                        edge from parent
                        node[above] {$q_m^{(1)}$}
                        node[below]  {}
                    }
                    child {
                        node[end, label=right:
                            {$ S_0 du $}] {}
                        edge from parent
                        node[above] {$q_u^{(1)}$}
                        node[below]  {}
                    }
                    edge from parent
                    node[above] {}
                    node[below]  {$q_d^{(0)}$}
            }
        child {
                node[bag] {$ S_0 m $}        
                     child {
                        node[end, label=right:
                            {$ S_0 md $}] {}
                        edge from parent
                        node[above] {}
                        node[below]  {$q_d^{(1)}$}
                    }
                    child {
                        node[end, label=right:
                            {$ S_0 mm $}] {}
                        edge from parent
                        node[above] {$q_m^{(1)}$}
                        node[below]  {}
                    }
                    child {
                        node[end, label=right:
                            {$ S_0 mu $}] {}
                        edge from parent
                        node[above] {$q_u^{(1)}$}
                        node[below]  {}
                    }
                    edge from parent
                    node[above] {$q_m^{(0)}$}
                    node[below]  {}
            }
            child {
                node[bag] {$ S_0 u $}        
                child {
                        node[end, label=right:
                            {$ S_0 ud $}] {}
                        edge from parent
                        node[above] {}
                        node[below]  {$q_d^{(1)}$}
                    }
                    child {
                        node[end, label=right:
                            {$ S_0 um $}] {}
                        edge from parent
                        node[above] {$q_m^{(1)}$}
                        node[below]  {}
                    }
                    child {
                        node[end, label=right:
                            {$ S_0 uu $}] {}
                        edge from parent
                        node[above] {$q_u^{(1)}$}
                        node[below]  {}
                    }
                edge from parent        
                    node[above] {$q_u^{(0)}$}
                    node[below]  {}
            };

            \draw (0,-5) -- (6,-5);
            \draw (0,-5.1) -- (0,-4.9)
                node at (0,-5.4) {0}
                node at (1.5,-4.8) {\small period $k=0$};
            \draw (3,-5.1) -- (3,-4.9)
                node at (3,-5.4) {$\Delta t$}
                node at (4.5,-4.8) {\small period $k=1$};
            \draw (6,-5.1) -- (6,-4.9)
                node at (6,-5.4) {2$\Delta t$};
        \end{tikzpicture}
        \caption{Stock price evolution\newline}
        \label{fig:stock-terminal}
     \end{subfigure}
     \hspace{2em}
     \begin{subfigure}[b]{0.4\textwidth}
        \centering
        \begin{tikzpicture}[grow=right, sloped]
        \node[bag] {$ S_0  $}
            child {
                node[bag] {$ S_0 d $}        
                    child {
                        node[end, label=right:
                            {$ S_0 dd $}] {}
                        edge from parent
                        node[above] {}
                        node[below]  {$q_d^{(1)}$}
                    }
                    child {
                        node[end, label=right:
                            {$ S_0 d e^{r\Delta t} $}] {}
                        edge from parent
                        node[above] {$q_m^{(1)}$}
                        node[below]  {}
                    }
                    child {
                        node[end, label=right:
                            {$ S_0 du $}] {}
                        edge from parent
                        node[above] {$q_u^{(1)}$}
                        node[below]  {}
                    }
                    edge from parent
                    node[above] {}
                    node[below]  {$q_d^{(0)}$}
            }
        child {
                node[bag] {$ S_0  e^{r\Delta t} $}        
                     child {
                        node[end, label=right:
                            {$ S_0  e^{2r\Delta t} $}] {}
                        edge from parent
                        node[above] {}
                        node[below]  {$q_d^{(1)}$}
                    }
                    child {
                        node[end, label=right:
                            {$ S_0  e^{2r\Delta t} $}] {}
                        edge from parent
                        node[above] {$q_m^{(1)}$}
                        node[below]  {}
                    }
                    child {
                        node[end, label=right:
                            {$ S_0  e^{2r\Delta t} $}] {}
                        edge from parent
                        node[above] {$q_u^{(1)}$}
                        node[below]  {}
                    }
                    edge from parent
                    node[above] {$q_m^{(0)}$}
                    node[below]  {}
            }
            child {
                node[bag] {$ S_0 u $}        
                child {
                        node[end, label=right:
                            {$ S_0 ud $}] {}
                        edge from parent
                        node[above] {}
                        node[below]  {$q_d^{(1)}$}
                    }
                    child {
                        node[end, label=right:
                            {$ S_0 u e^{r\Delta t} $}] {}
                        edge from parent
                        node[above] {$q_m^{(1)}$}
                        node[below]  {}
                    }
                    child {
                        node[end, label=right:
                            {$ S_0 uu $}] {}
                        edge from parent
                        node[above] {$q_u^{(1)}$}
                        node[below]  {}
                    }
                edge from parent        
                    node[above] {$q_u^{(0)}$}
                    node[below]  {}
            };
            \draw (0,-5) -- (6,-5);
            \draw (0,-5.1) -- (0,-4.9)
                node at (0,-5.4) {0}
                node at (1.5,-4.8) {\small period $k=0$};
            \draw (3,-5.1) -- (3,-4.9)
                node at (3,-5.4) {$\Delta t$}
                node at (4.5,-4.8) {\small period $k=1$};
            \draw (6,-5.1) -- (6,-4.9)
                node at (6,-5.4) {2$\Delta t$};
        \end{tikzpicture}
        \caption{Early delivery on middle paths with carried interest until maturity}
        \label{fig:stock-random}
     \end{subfigure}
     \caption{Stock price trees (no dividends).}
     \label{fig:stocks}
\end{figure}

The one-period trinomial model is easily iterated into a tree with constant ex-dividend stock price movement factors $u > m > d > 0$ over $N$ time periods of equal length $\Delta t = T/N$ , where $T$ represents some time horizon. For ease of notation, we adopt zero-based indexing $k = 0, 1, \dots, N-1$ of the successive time periods defined by the $N+1$ time points $t = 0, \Delta t, \dots, k\Delta t, \dots, (N-1)\Delta t, N\Delta t = T$. Figure \ref{fig:stock-terminal} shows a standard trinomial tree modelling the evolution of the stock price over $N=2$ periods.  Now consider the case with random expiry, where we interpret taking the middle path as random expiry.  A two-period trinomial tree modelling this situation is given in Figure \ref{fig:stock-random}.  Observe how interest is carried to maturity after the stock is delivered early, so that the nine terminal nodes correspond to the payoff with carried interest.

This approach effectively embeds an RE payoff into a standard trinomial tree. It turns out that Figure~\ref{fig:stock-random} describes the price evolution of an RE-ZSC option on a non-dividend-paying stock, as shown by Proposition~\ref{prop:stock-random-expiry}, given below.  A more general illustration for an arbitrary payoff $f(S)$ is provided later in Figure~\ref{fig:RE-option}.

In the context of the multi-period trinomial tree, let $S_k$ be the random variable representing the ex-dividend stock price at the beginning of period $k = 0, 1, \dots, N-1$, and let $S_N$ be the random variable representing the ex-dividend terminal stock price (as illustrated in Figure \ref{fig:stock-terminal} for $N=2$). It is worth emphasizing that  $S_k$ is \emph{not} a tradable asset price in the sense of Section \ref{sec:theoretical-background} because it differs from the stock payoff by the amount of accumulated dividends, e.g.,\ for $k=2$ the up-up stock payoff is $S_0 u^2 e^{2y\Delta t}$ but the ex-dividend stock price is $S_2 = S_0 u^2$.

Let $\tau\in\{0, 1, \dots, N\}$ be the random variable corresponding to the period when the middle path is first taken, with the convention that $\tau = N$ if the middle path is never taken; mathematically, $\tau := \min \{0 \leq k\leq N-1: S_{k+1}=m S_k \}$ with $\min \emptyset = N$. We interpret $\tau$ as the period when random expiry occurs, i.e.,\ the time of random expiry is $\tau\Delta t$. We shall refer to $\tau$ as the random expiry throughout the paper.

We now describe the stock price evolution and risk-neutral probabilities on the trinomial tree. We first consider the general situation, where these probabilities can be path-dependent, and then specialize to the so-called homogeneous case with path-independent risk-neutral probabilities. The latter is the situation illustrated in Figure~\ref{fig:stocks} and is the basis for most of our results.

\subsubsection{General trinomial tree}\label{sec: general trinomial tree}

In this section, we consider the trinomial tree in its most general form. Every node in the tree can be identified with the path taken to get there. Thus, any node $\varsigma$ at time period $k=0,1,\dots, N$ will be denoted by an element of $\{u,m,d\}^k$ with the convention that $\{u,m,d\}^0=\emptyset$ denotes the root node. For each node $\varsigma\in\{u,m,d\}^k$, $k=0,1,\dots, N-1$, we let $q_u^{(k,\varsigma)}, q_m^{(k,\varsigma)}, q_d^{(k,\varsigma)}$ be the risk-neutral probabilities of going up, middle, or down, respectively.  In this setup, there are $3^N$ states of the world $\omega_i$, each corresponding to a full path $(\epsilon_{i,1}, \dots, \epsilon_{i,N})\in\{u,m,d\}^N$ of $N$ trinomial moves with compound risk-neutral probability $\Q(\{\omega_i\}) = q_{\epsilon_{i,1}}^{(0,\emptyset)} q_{\epsilon_{i,2}}^{(1, \epsilon_{i,1})} \cdots q_{\epsilon_{i,N}}^{(N-1, \epsilon_{i,1},\dots,\epsilon_{i,N-1})}$. Note that each state of the world corresponds to a terminal node. Let $\varSigma_1,\varSigma_2,\dots,\varSigma_N$ be the random sequence of nodes visited. It is easily checked that this sequence forms a Markov process. On the other hand, the random sequence of observed  ex-dividend stock prices $(S_k)_{0\leq k\leq N}$ does not, in general, form a Markov process.

Throughout, we take $m=e^{(r-y)\Delta t}$, which, as discussed in Section \ref{sec:one-period-trinomial}, means that we may select $q_m^{(k,\varsigma)}\in(0,1)$ freely at every period $k$ and node $\varsigma\in\{u,m,d\}^k$ without introducing arbitrage.  This will allow us to choose any distribution for $\tau$ so long as $\Q(\tau = 0) \in(0,1)$ and $\Q(\tau = k\ |\ \varSigma_k=\varsigma)\in(0,1)$ for $k \geq 1$ and $\varsigma\in\{u,d\}^k$. We may then calibrate the risk-neutral probabilities by
\begin{equation}    \label{eq:RN-trinomial-Nstep-drift-general}
    \begin{cases}
        \displaystyle
        q_u^{(k,\varsigma)} = \frac{e^{(r-y)\Delta t}-d}{u-d}(1 - q_m^{(k,\varsigma)}) \\
        q_m^{(k,\varsigma)} = \Q(\tau = k\ |\ \varSigma_k = \varsigma) \\
        \displaystyle
        q_d^{(k,\varsigma)} =  \frac{u-e^{(r-y)\Delta t}}{u-d}(1 - q_m^{(k,\varsigma)})
    \end{cases}
    ,\quad k = 0,1,\dots,N-1,\ \varsigma\in\{u,d\}^k.
\end{equation}
Note that, in \eqref{eq:RN-trinomial-Nstep-drift-general}, we have $\varsigma\in\{u,d\}^k$, which implies that $[\varSigma_k = \varsigma]\subset[\tau\ge k]$. Note further that $\bigcup_{\varsigma\in\{u,d\}^k}[\varSigma_k = \varsigma]=[\tau\ge k]$.
It remains to calibrate the risk-neutral probabilities for nodes after we have taken a middle path, i.e., for $\varsigma\in\{u,m,d\}^k\setminus\{u,d\}^k$, where $\setminus$ denotes the set difference. For such $\varsigma$, we can choose $q_m^{(k,\varsigma)}\in(0,1)$ arbitrarily and then use \eqref{eq:RN-trinomial-Nstep-drift-general} to calibrate $q_u^{(k,\varsigma)}$ and $q_d^{(k,\varsigma)}$.

\subsubsection{Homogeneous trinomial tree}

In this section we consider a specialization of the model described in Section \ref{sec: general trinomial tree}, to the case where the risk-neutral probabilities are no longer allowed to depend on the given node, but only on the period number. We call this the homogeneous trinomial tree model; it is the main focus of this paper. In this case, for each period, $k=0,1,\dots, N-1$, we let $q_u^{(k)}$, $q_m^{(k)}$, and $q_d^{(k)}$ be the risk-neutral probabilities of going up, middle, or down, respectively. In this setup, there are 
$3^N$ states of the world $\omega_i$, each corresponding to a path $(\epsilon_{i,1}, \dots, \epsilon_{i,N})\in\{u,m,d\}^N$ of $N$ trinomial moves with compound risk-neutral probability $\Q(\{\omega_i\}) = q_{\epsilon_{i,1}}^{(0)} \cdots q_{\epsilon_{i,N}}^{(N-1)}$.  This standard tree construction ensures that, under $\Q$, the random sequence of ex-dividend stock prices $(S_k)_{0\leq k\leq N}$ forms a Markov process and that the random variables $S_1/S_0,S_2/S_1,\dots,S_N/S_{N-1}$ are independent.

As with the general trinomial tree discussed in Section \ref{sec: general trinomial tree}, we take $m=e^{(r-y)\Delta t}$ and may then 
select $q_m^{(k)}\in(0,1)$ freely at every period $k$.  In other words, any distribution for the random expiry $\tau$\ is risk-neutral, so long as $\Q(\tau = k) \in (0,1)$ is satisfied for each $k=0,1,\dots,N$. Once we select such a distribution for $\tau$, we may calibrate the risk-neutral probabilities as
\begin{equation}    \label{eq:RN-homogeneous-trinomial-Nstep-drift}
    \begin{cases}
        \displaystyle
        q_u^{(k)} = \frac{e^{(r-y)\Delta t}-d}{u-d}(1 - q_m^{(k)}) \\
        q_m^{(k)} = \Q(\tau =k\ |\ \tau \geq k  ) \\
        \displaystyle
        q_d^{(k)} =  \frac{u-e^{(r-y)\Delta t}}{u-d}(1 - q_m^{(k)})
    \end{cases}
    , \quad k = 0,1,\dots,N-1.
\end{equation}
For example, if we assume constant $q_m^{(k)} = p$ for $k = 0,1,\dots,N-1$, then $\tau$ has a finite geometric distribution with probability mass function
\begin{eqnarray}\label{eq: finite geo}
    \Q( \tau = k ) = p(1-p)^k, \; k=0,1,\dots,N-1; \qquad \Q( \tau = N ) = (1-p)^N.
\end{eqnarray}

We note that, in principle, we can allow for the case where $\Q(\tau = k)=0$ or $=1$ for some $k$. However, for this to work, we need to modify not just the measure $\Q$, but the underlying tree. Specifically, if this equals $1$ for some $k$, then we can dispense with the trinomial tree entirely, and use a classical binomial tree model with $k$ periods. On the other hand, if this is equal to $0$ for some $k$, then we need to assume that at location $k$ our tree has only two edges instead of three, and that we are \emph{locally} operating in a two-state complete market economy. At any location $k$ with $\Q(\tau = k)\in(0,1)$, we still assume that we have three edges. With these minor changes, the results in this paper will still hold. However, in order to avoid unnecessary complications, we assume that $\Q(\tau = k) \in (0,1)$ throughout. We note that similar modifications can also be introduced in the general trinomial tree case.

\section{Pricing random-expiry options}     \label{sec:RE-pricing}

In this section, we give our main results for pricing RE options.

\begin{theorem}\label{thm:main}
In the $N$-period general trinomial tree with constant interest rate $r\ge0$, constant dividend yield $y\ge0$, constant ex-dividend movement factors $ u > m = e^{(r-y)\Delta t} > d > 0$, and risk-neutral probabilities given by \eqref{eq:RN-trinomial-Nstep-drift-general}, we have:
\begin{enumerate}[label=\normalfont{\bf (\alph*)},leftmargin=*]
    \item 
    
    A no-arbitrage price of an RE option with payoff $f(S_\tau)$  paid at random expiry $\tau\in\{0, 1, \dots, N\}$ is given by
\[
    V_0 = \E^{\Q}\!\big(e^{-r\tau\Delta t} f(S_\tau) \big) = \sum_{k=0}^N e^{-rk\Delta t} \E^{\Q}(f(S_k)\ |\ \tau =k) \Q(\tau = k).
\]
    \item The stock price process with dividends discounted at the risk-free rate $\hat S_k = e^{(y-r)k\Delta t}S_k$, $k=0,1,\dots,N$, is a $\Q$-martingale.
\end{enumerate}
In the $N$-period homogeneous tree with path-independent probabilities given by \eqref{eq:RN-homogeneous-trinomial-Nstep-drift}, we further have:
\begin{enumerate}[label=\normalfont{\bf (\alph*)},leftmargin=*,start=3]
    \item Conditionally upon $[\tau \geq l]$ for any $0\leq  l \leq N-1$, $\tau$ is independent from $(S_k)_{0\leq k\leq l}$.
    \item The joint distribution of $\tau$ and the stopped process $(S_{k\wedge\tau})_{0\leq k\leq N}$ may be rewritten, under the risk-neutral measure $\Q$, as
    \[
        \big(\tau, S_{1\wedge\tau}, \dots, S_{N\wedge\tau}\big) \eqd \left(\tau, S_0 \exp\!\left(\sum_{l=1}^{1\wedge\tau} X_{ l }\right),\dots,S_0 \exp\!\left(\sum_{l=1}^{N\wedge\tau} X_{ l }\right)\right),
    \]

    where $(X_{ l })_{1\leq  l \leq N}$ is a sequence of mutually independent random variables that is independent of $\tau$ and satisfies
    \begin{equation}    \label{eq:RN-binomial}
    \begin{dcases}
        \Q(e^{X_{ l }} = u)  = \frac{e^{(r-y)\Delta t}-d}{u-d} \\
        \Q(e^{X_{ l }} = d) = 1-\Q(e^{X_{ l }} = u) = \frac{u-e^{(r-y)\Delta t}}{u-d}
    \end{dcases}
    , \quad l = 1,2,\dots, N.
    \end{equation}
    \end{enumerate}
\end{theorem}
\begin{proof}
    See Appendix \ref{app:proof-th1}.
\end{proof}

\begin{remark}
    In the above, for $1\le l\le\tau$, $X_l$ is a random variable that tells us if the stock went up or down at step $l-1$. 
    Note that, by the definition of $\tau$, at step $l-1$ we could not have taken the middle path. Here, the indexing is chosen to simplify the notation.
\end{remark}

\begin{remark}  \label{rem:binomial-pricing}
Note that $\tau=k$ means that we never take the middle path in periods $0$ to $k-1$, but that we take it in period $k$. Thus, we can calculate the conditional expectation in Part~(a) using a $k$-step \emph{binomial} tree covering periods $0$ to $k-1$ with constant ex-dividend movement factors $u,d$, and constant risk-neutral probabilities $q_u = \frac{e^{(r-y)\Delta t}-d}{u-d}, q_d = 1 - q_u$. This binomial tree has $2^k$ terminal nodes corresponding to the possible values of $S_k$ and we can calculate $\E^{\Q}(f(S_k)\ |\ \tau =k)$ by standard backward induction. However, this approach would require pricing $N$ fixed-expiry options, each on a seperate tree. More efficient numerical methods are discussed in Section~\ref{sec:numerical}.
\end{remark}

From the proof of Theorem \ref{thm:main} it is readily seen that Parts~(c) and (d) crucially depend on the assumption that the risk-neutral probabilities are path-independent, and would thus not hold for general trinomial trees.  In particular, if risk-neutral probabilities were path-dependent, then $\tau$ would be dependent on the stock price process, and the random variables $X_1,X_2,\dots,X_N$ would not be independent. It is worth emphasizing that Part~(c) implies that the event triggering early expiry cannot depend on the stock price. This excludes situations where $\tau$ is optimally determined by the stock price as is the case with American options, or even correlated situations such as credit derivatives. Thus, the homogeneous case corresponds to applications where the event triggering early expiry does not depend on the stock price.

\begin{assumption}
Throughout the remainder of this paper, we assume that the trinomial tree is homogeneous.    
\end{assumption}

The following proposition stems from Theorem~\ref{thm:main}(a) applied to $f(S)=S$.

\begin{proposition} \label{prop:stock-random-expiry}
    The no-arbitrage price of the random-expiry zero-strike call (RE-ZSC) is
\begin{equation}    \label{eq:stock-random-expiry-price}
    V_0^{\text{\em RE-ZSC}} = \E^{\Q}(e^{-r\tau\Delta t} S_\tau) = S_0 \E^{\Q}(e^{-y\tau\Delta t}).
\end{equation}
\end{proposition}
\begin{proof}
    See Appendix~\ref{app:proof-th1}.
\end{proof}

Note that $\E^{\Q}(e^{-y\tau\Delta t}) $ is the moment-generating function of $\tau$ evaluated at point $-y\Delta t$. In addition, it should be mentioned that the result in Proposition~\ref{prop:stock-random-expiry} is consistent with financial intuition. First, note that, by Remark~\ref{rem:binomial-pricing}, $e^{-rk\Delta t} \E^{\Q}(S_k\ |\ \tau =k)$ is the risk-neutral price of a fixed-expiry ZSC, which may be calculated using a $k$-step binomial tree and must equal $S_0 e^{-yk\Delta t}$. The second intuition is for the case of a non-dividend-paying stock (where $y=0$), for which Proposition~\ref{prop:stock-random-expiry} implies that the risk-neutral price is simply $S_0$. This is in agreement with the simple fact that such a claim is perfectly hedged by buying a share of the stock at time $0$ for $S_0$ and delivering it at time $\tau \Delta t$.  While here, this static hedging strategy perfectly replicates the claim, in general, perfect replication of an RE option is not possible just by trading the stock. This limitation also applies to an RE-ZSC option on a dividend-paying stock.

\begin{proposition}   \label{cor:RE-price-range}
    The no-arbitrage price of the RE option with payoff $f(S_\tau)$ is in the range
\[
        V_0 \in \underset{k=0,1,\dots,N}{(\min,\max)}\  e^{-rk\Delta t} \E^{\Q}(f(S_{k})\ |\ \tau =k).
    \]  
If, in addition, $e^{-rk\Delta t} \E^{\Q}(f(S_k)\ |\ \tau =k)$ is monotonic in $k$, then the no-arbitrage price of the RE option is in the range
    \[
        V_0 \in \Big(\min\!\big(f(S_0), e^{-rT} \E^{\Q}(f(S_N)\ |\ \tau=N)\big),\ \max\!\big(f(S_0), e^{-rT} \E^{\Q}(f(S_N) |\ \tau=N)\big) \Big).
    \]
\end{proposition}
\begin{proof}
    See Appendix~\ref{app:proof-th1}.
\end{proof}

In many practical situations we may be interested in optimizing the price not over all possible distributions of $\tau$, but only over those that belong to some relevant class. 
In this case, the space of distributions for $\tau$ could be constrained in  complicated ways and determining the price range of the RE option may be nontrivial.  However, if the constraints are linear inequalities of the form $ \underline p \leq \sum_k \alpha_k \Q(\tau=k) \leq \bar p$, we have a linear programming problem, whose solution, if feasible, is known to reside on the boundary \citep[Ch.~2]{bertsimas-tsitsiklis:1997}. For example, if $\tau$ has a finite geometric distribution with parameter $p\in[ \underline p, \bar p]\subset(0,1)$ and the fixed-expiry option price $e^{-rk\Delta t} \E^{\Q}(S_k\ |\ \tau =k)$ is monotonic, then the price range is attained at the boundaries $p=\underline p$ and $p = \bar p$.

We conclude this section by noting that, {\em mutatis mutandi}, all of our results can be straightforwardly extended to path-dependent payoffs $f(S_0, S_1, \dots, S_\tau)$.  Examples include, e.g., Asian, barrier, forward-start, lookback, and cliquet options---see, e.g., \citet[ch.12]{bossu:2012} or \citet[ch.1]{bossu:2014}. It is worth emphasizing that the pricing of path-dependent RE options can be done on homogeneous trees and does not, necessarily, require one to use the general trinomial tree model with path-dependent risk-neutral probabilities.

\section{Continuous-time limits} \label{sec:continuous}

Fix a time horizon $T$ and, for simplicity, assume that $T$ is an integer. Consider a sequence of trinomial trees of the form \ref{fig:stock-random}, where the $n$th tree has $Tn$ time periods. Thus, we have $n$ periods per unit time, each of length $\Delta t = 1/n$. Let $\tau_n$ be the random expiry for the $n$th tree. Assume that $\tau_n$ is supported on $\{0,1,\dots,nT\}$ for each $n$ and that there exists a random variable $\tau_\infty$ with support contained in $[0,T]$ satisfying
\begin{eqnarray}\label{eq: assump on tau n}
       \frac{\tau_n}{n}\conp\tau_\infty.
\end{eqnarray}
Here, and throughout this section, all limits in distribution, in probability, and with probability $1$ are taken under the risk-neutral measure. To make the asymptotics work, we need to select specific forms for the parameters $u$, $m$, and $d$. Toward this end, fix $\sigma>0$ and, henceforth, assume that $n>\sigma^2$. We take
\begin{equation*}    
    \left\{\begin{aligned}
        u_n & = e^{(r-y)/n}\left(1 + \sigma/\sqrt n\right) \\
        m_n & = e^{(r-y)/n} \\
        d_n & = e^{(r-y)/n}\left(1 - \sigma/\sqrt n\right) 
    \end{aligned}\right.
\end{equation*}
to be, respectively, $u$, $m$, and $d$ in the $n$th tree.
Using \eqref{eq:RN-homogeneous-trinomial-Nstep-drift}, we find that, for the $n$th tree, the risk-neutral probabilities at period $k$ reduce to
\begin{equation*} 
    \begin{cases}
        \displaystyle
        q_{u,n}^{(k)} = \frac{e^{(r-y)/n}-d_n}{u_n-d_n}(1 - q_{m,n}^{(k)})=\frac{1}{2}(1 - q_{m,n}^{(k)})  \\
        q_{m,n}^{(k)} = \Q(\tau_n =k\ |\ \tau_n \geq k ) \\
        \displaystyle
        q_{d,n}^{(k)} =  \frac{u_n-e^{(r-y)/n}}{u_n-d_n}(1 - q_{m,n}^{(k)})=\frac{1}{2}(1 - q_{m,n}^{(k)})
    \end{cases}
    , \quad k = 0,1,\dots,Tn-1.
\end{equation*}

We embed the $n$th tree in a continuous time model as follows. We take $S_n(t)$ for $t\in[0,1]$ to be the ex-dividend stock price on the $n$th tree 
at period $\floor{nt}$. Next, let $\tau_n^*(t) = \frac{\floor{nt}\wedge\tau_n}{n}$ and note that Theorem \ref{thm:main}(d) implies that
$$
S_n^*(t):=S_n(\tau_n^*(t)) \eqd S(0) \exp\left\{\sum_{k=1}^{\floor{ nt}\wedge\tau_n} X_{k,n}\right\} , \qquad t\in[0,T],
$$
where $S(0)$ is the initial price of the stock and $(X_{k,n})_{k=1,2,\dots,Tn}$ are iid random variables, independent of $\tau_n$, which satisfy 
$$
\Q(e^{X_{ k,n}} = u_n)  =  \Q(e^{X_{k,n }} = d_n) =\frac{1}{2}.
$$
Next, let $Y_1,Y_2,\dots$ be iid random variables, independent of $\tau_n$ for each $n$, such that
$$
\Q(Y_{k} = 1)  =  \Q(Y_{k} = -1) =\frac{1}{2}
$$
and note that
$$
X_{k,n} \eqd \frac{r-y}{n} + \log(1+Y_k \sigma/\sqrt n).
$$
Furthermore, in light of our various assumptions of independence, we have
$$
S_n^*(t)=S_n(\tau_n^*(t)) \eqd S(0) \exp\left\{\frac{r-y}{n}\left(\floor{nt}\wedge\tau_n\right) + \sum_{k=1}^{\floor{ nt}\wedge\tau_n} \log(1+Y_k \sigma/\sqrt n) \right\} , \qquad t\in[0,T].
$$

Next, we introduce our limiting process. Let $\left\{W_t:t\in[0,T]\right\}$ be standard Brownian motion independent of $\tau$ and set
\[
    S_\infty(t) = S(0) e^{ \left(r-y - \frac12 \sigma^2 \right)t + \sigma W_t} ,\qquad t\in[0,T].
\]
Let $\tau^*_\infty(t)=\tau_\infty\wedge t$ and note that
\[
    S_\infty^*(t):=S_\infty(\tau^*_\infty(t)) = S(0) e^{ \left(r-y - \frac12 \sigma^2 \right) (\tau_\infty\wedge t) + \sigma W_{\tau_\infty\wedge t}},\qquad t\in[0,T].
\]

\begin{lemma} \label{lem:sup-to-zero}
We have $\tau_n/n\conp\tau_\infty$ if and only if
\begin{eqnarray*}
    \sup_{t\in[0,T]}\left|\tau_n^*(t) -\tau_\infty^*(t)\right| = \sup_{t\in[0,T]}\left|\frac{\tau_n\wedge\floor{ nt}}{n} -\tau_\infty\wedge t\right| \conp 0.
\end{eqnarray*}
\end{lemma}
\begin{proof}
    See Appendix~\ref{app:proof-th2-3}.
\end{proof}

Let $D([0,T])$ be the space of c\`adl\`ag functions (i.e.,\ functions that are right continuous with left limits) on $[0,T]$ equipped with the Skorohod topology \citep[see][Ch.\ 3]{Billingsley:1999}. Note that $\{S_n^*(t):t\in[0,T]\}\in D([0,T])$ for $n=1,2,\dots,\infty$, that $S^*_n(T)=S_n(\tau_n/n)$ for $n=1,2,\dots$, and that $S^*_\infty(T)=S_\infty(\tau_\infty)$.

\begin{theorem}\label{thrm: limits}
\textbf{\em (a)}~We have
$$
\{S^*_n(t): t\in[0,T]\} \cond \{S^*_\infty(t) :t\in[0,T]\} \mbox{ on }D([0,T])
$$
and for any continuous function $f:\mathbb R\to\mathbb R$
$$
\{e^{-r\tau_n/n}f(S^*_n(t)): t\in[0,T]\} \cond \{e^{-r\tau}f(S^*_\infty(t)) :t\in[0,T]\} \mbox{ on }D([0,T]).
$$
\textbf{\em (b)}~If $f:\mathbb R\to\mathbb R$ is a bounded and continuous function, then
$$
\lim_{n\to\infty} \E^{\Q} \left[e^{-r\tau_n/n} f(S_n(\tau_n/n))\right] = \E^{\Q} \left[e^{-r\tau_\infty} f(S_{\infty}(\tau_\infty))\right].
$$
\end{theorem}
\begin{proof}
    See Appendix~\ref{app:proof-th2-3}.
\end{proof}

Note that $S_n(\tau_n/n)$ is the ex-dividend price at time $\lfloor n\tau_n/n\rfloor=\tau_n$. Thus, $\E^{\Q} \left[e^{-r\tau_n/n} f(S_n(\tau_n/n))\right]$ corresponds exactly to the expectation in Theorem \ref{thm:main}(a). Theorem~\ref{thrm: limits}(b) implies that we can approximate expectations of the form $\E^{\Q} \left[e^{-r\tau_\infty} f(S_{\infty}(\tau_\infty))\right]$ by pricing RE options with payoff function $f$ on a trinomial tree. This is useful for pricing RE options in continuous time. Moreover, such expectations can arise in other contexts, e.g., when working with functions of subordinated Brownian motion, which is an important class of infinitely divisible distributions. Some examples that arise in financial applications include: variance gamma distributions (\cite{Madan:Carr:Chang:1998}), normal inverse Gaussian distributions (\cite{Barndorff-Nielsen:1997}, \cite{Eriksson:2009}), and normal tempered stable distributions (\cite{Grabchak:2016}, \cite{Azzone:2022}, \cite{Sabino:2023}). In such situations, the support of $\tau_\infty$ is typically unbounded, but we can always approximate by a random variable whose support is contained in $[0,T]$ for some large $T$. 

We conclude this section by giving an approach for constructing random variables $\tau_n$ that satisfy \eqref{eq: assump on tau n}. 
Let $\tau_\infty$ be any random variable on $[0,T]$ and let $\tau_n=\floor {n\tau_\infty}$. 
It is easily checked that $\tau_n/n\to\tau_\infty$ with probability $1$, which is formally stronger than \eqref{eq: assump on tau n}. When the distribution of $\tau_\infty$ is continuous, we have $\Q(\tau_\infty=T)=0$ and then $\floor {n\tau_\infty}\le nT-1$ with probability $1$. In this case, we can take $\tau_n=\floor {n\tau_\infty+1}$ to retain the possibility of no early expiry on the $n$th tree. 

We now give an example in the context of the finite geometric distribution, which has $\Q(\tau_\infty = T) > 0$. 
Fix $\lambda>0$, let $\pi=e^{-\lambda T}$, and note that $\pi\in(0,1)$. Assume that the distribution of $\tau_\infty$ is given by
$$
(1-\pi) \frac{\lambda}{1-e^{-\lambda T}} e^{-\lambda x}\I_{[0<x<T]}\rd x + \pi \delta_T(\rd x) = \lambda e^{-\lambda x}\I_{[0<x<T]}\rd x + e^{-\lambda T} \delta_T(\rd x),
$$
where $\delta_T$ denotes the Dirac point mass at $T$. This is a mixture such that, with probability $\pi$ we have $\tau_\infty=T$ and with probability $1-\pi$ we have $\tau\eqd X$, where $X$ is a continuous random variable with density $f(x) = \frac{\lambda}{1-e^{-\lambda T}}e^{-\lambda x}$, $0<x<T$. It is easy to check that $\tau_n=\floor{n\tau}$ has a finite geometric distribution, as given in \eqref{eq: finite geo}, with $N=Tn$ and $p=1-e^{-\lambda/n}$. 

\section{Numerical implementation}      \label{sec:numerical}

In this section we introduce and compare three algorithms for pricing RE options. A Python implementation is given in Appendix~\ref{app:code}.  The full code, which contains additional programs to reproduce the graphs, can be found on GitHub. The algorithms are summarized in Table~\ref{tab:algo-list} and work as follows.

\begin{figure}[t]
    \centering
    \begin{tikzpicture}[grow=right, sloped, every node/.style={align=center, sibling distance=5cm}]
    \node[align=right] {$ V(0, S_0)  $}
        child {
            node[] {$ V(1, S_0 d) $}        
                child {
                    node[end, label=right:
                        {$ f(S_0 dd) $}] {}
                    edge from parent
                    node[above] {}
                    node[below]  {$q_d^{(1)}$}
                }
                child {
                    node[end, label=right:
                        {$ f(S_0 d) e^{r\Delta t} $}] {}
                    edge from parent
                    node[above] {$q_m^{(1)}$}
                    node[below]  {}
                }
                child {
                    node[end, label=right:
                        {$ f(S_0 du) $}] {}
                    edge from parent
                    node[above] {$q_u^{(1)}$}
                    node[below]  {}
                }
                edge from parent
                node[above] {}
                node[below]  {$q_d^{(0)}$}
        }
    child {
            node[] {$ f(S_0)  e^{r\Delta t} $}        
                 child {
                    node[end, label=right:
                        {$ f(S_0)  e^{2r\Delta t} $}] {}
                    edge from parent
                    node[above] {}
                    node[below]  {$q_d^{(1)}$}
                }
                child {
                    node[end, label=right:
                        {$ f(S_0)  e^{2r\Delta t} $}] {}
                    edge from parent
                    node[above] {$q_m^{(1)}$}
                    node[below]  {}
                }
                child {
                    node[end, label=right:
                        {$ f(S_0)  e^{2r\Delta t} $}] {}
                    edge from parent
                    node[above] {$q_u^{(1)}$}
                    node[below]  {}
                }
                edge from parent
                node[above] {$q_m^{(0)}$}
                node[below]  {}
        }
        child {
            node[] {$ V(1, S_0 u) $}        
            child {
                    node[end, label=right:
                        {$ f(S_0 ud) $}] {}
                    edge from parent
                    node[above] {}
                    node[below]  {$q_d^{(1)}$}
                }
                child {
                    node[end, label=right:
                        {$ f(S_0 u) e^{r\Delta t} $}] {}
                    edge from parent
                    node[above] {$q_m^{(1)}$}
                    node[below]  {}
                }
                child {
                    node[end, label=right:
                        {$ f(S_0 uu) $}] {}
                    edge from parent
                    node[above] {$q_u^{(1)}$}
                    node[below]  {}
                }
            edge from parent        
                node[above] {$q_u^{(0)}$}
                node[below]  {}
        };
        \draw (0,-5) -- (6,-5);
        \draw (0,-5.1) -- (0,-4.9)
            node at (0,-5.4) {0}
            node at (1.5,-4.8) {\small period $k=0$};
        \draw (3,-5.1) -- (3,-4.9)
            node at (3,-5.4) {$\Delta t$}
            node at (4.5,-4.8) {\small period $k=1$};
        \draw (6,-5.1) -- (6,-4.9)
            node at (6,-5.4) {2$\Delta t$};
    \end{tikzpicture}
    \caption{Option value for general RE payoff $f(S)$. $V(k,S_k)$ is the value at time $k\Delta t$ and stock price node $S_k$.}
    \label{fig:RE-option}
\end{figure}
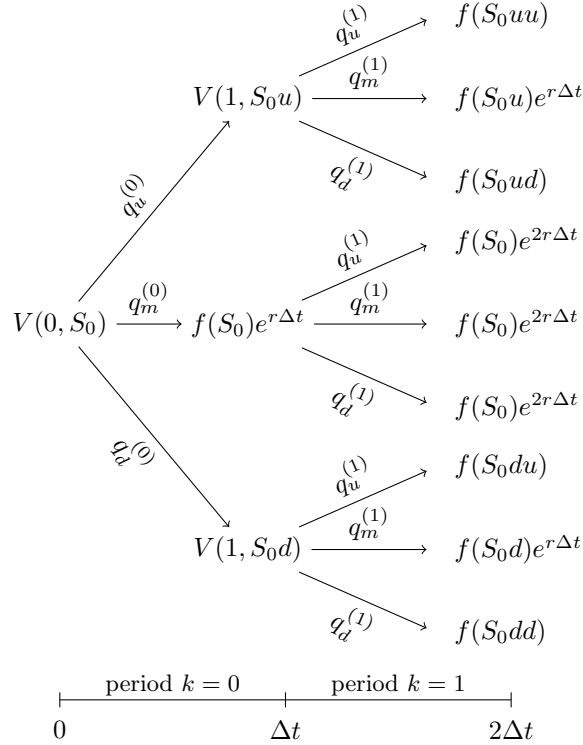

Algorithm~1, which serves as our baseline, creates two classical trinomial trees as illustrated in Figures~\ref{fig:stock-terminal} and~\ref{fig:RE-option}, respectively, for the stock price and  for the RE option price. The algorithm populates the option tree with terminal payoff values as well as early payoffs on all middle nodes and their descendants with carried interest; we refer to this as \emph{ad hoc node initialization}. Pricing of the remaining nodes is performed using standard backward induction.

Algorithms~2 and 3 use a modified binomial tree methodology based on the fact that the stopped process $(S_{k\wedge\tau})$ is, almost surely, a binomial random walk (Theorem~\ref{thm:main}(d)).  Algorithm~2 is based on a classic non-recombining binomial tree with  exponential growth, while Algorithm~3 is based on a recombining binomial tree with linear growth. These algorithms correspond to the stopped trinomial tree illustrated in Figure~\ref{fig:trinomial-early}, with the caveat that the middle path (which is taken only once) is considered virtual and not coded in memory. Both algorithms use a modified backward pricing equation, whereby each binomial node leads to three outcomes: the up and down child nodes in the binomial tree and a virtual middle path that is not coded.  Specifically, at the beginning of each period $k=N-1,N-2,\dots,0$, the option value is given by the trinomial risk-neutral equation
\begin{equation}    \label{eq:trinomial-pricing}
    V(k, S_k) = e^{-r\Delta t} q_u^{(k)} V(k+1, S_k u) + q_m^{(k)} f(S_k) + e^{-r\Delta t} q_d^{(k)} V(k+1, S_k d),
\end{equation}
where $V(k,S_k)$ is the option value at time $k\Delta t$ and stock price $S_k$, $f(S_k)$ is the payoff at time $k\Delta t$ when $\tau = k$, and $V(k+1,S_k u)$ and $V(k+1,S_k d)$ denote, respectively, the option value at the up and down child nodes. 

Algorithm 3 has the advantage of linear growth; however, it cannot be easily generalized to more complicated situations. In particular, it would not be suitable for the path-dependent risk-neutral probabilities in the general tree discussed in Section~\ref{sec: general trinomial tree} or even for path-dependent payoffs discussed at the end of Section~\ref{sec:trinomial}. On the other hand, Algorithms 1 and 2 could be easily modified to work in these cases.

We tested our Python implementation on the four RE options listed in Table~\ref{tab:payoff-list}.  All tests were performed using the following  parameters: $\Delta t = T/N$, $m = e^{(r-y)\Delta t}$,$u,d = m e^{\pm\sigma\sqrt{\Delta t}}$, and path-independent probabilities $q_m = \lambda \Delta t$, with $q_u, q_d$ given by \eqref{eq:RN-homogeneous-trinomial-Nstep-drift}. Here, $\sigma,\lambda>0$ are parameters, which we refer to as the volatility and the expiry intensity, respectively. 

\begin{figure}[ht]
    \centering
    \includegraphics[width=0.5\textwidth]{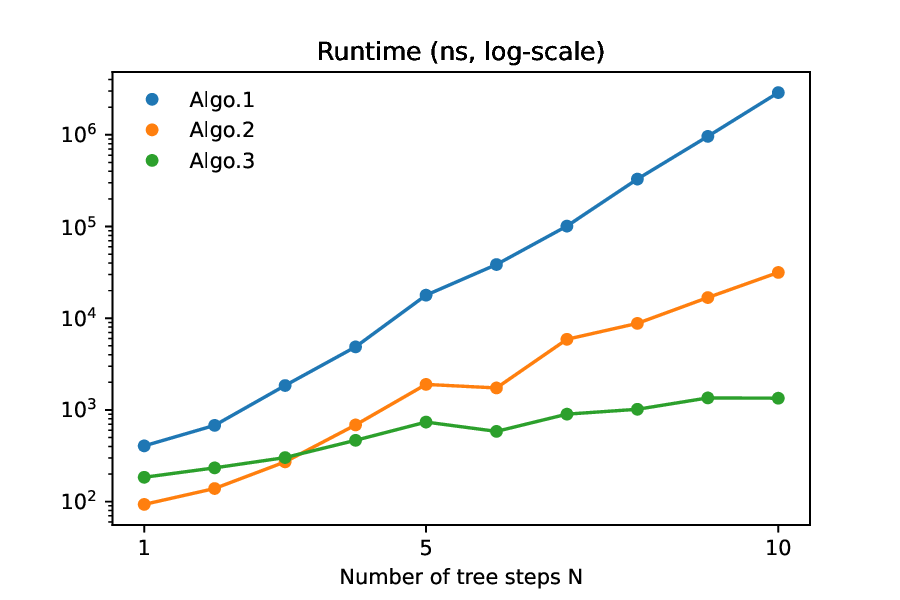}
  \caption{Execution time in nanoseconds (ns) for the RE call as a function of the number of time periods $N$.}  \label{fig:runtime-against-N}
\end{figure}

Our numerical tests were performed on a 13th~Gen Intel\textregistered~Core\texttrademark\ i7-13850HX 2.10~GHz processor. First, we checked that the algorithms give consistent results. We obtained identical numerical values up to the fourth decimal place across 1000 uniformly randomized parameterizations. Next, we examined execution times, which are reported in Figure~\ref{fig:runtime-against-N}.  For each algorithm we report the time averaged over 10 runs with identical parameters (see Table~\ref{tab:numerical-params}), as the number of time periods increases from 1 to 10. We refer the reader to our code on GitHub for additional details. Predictably, Algorithm~3 is the fastest thanks to its linear growth. For this reason, we only considered this algorithm for the remaining numerical experiments.

\begin{table}[htbp]
    \caption{Algorithms for RE option pricing}
    \label{tab:algo-list}
    \centering
    \begin{tabular}{|c|l|p{0.55\linewidth}|}
        \hline
        \textbf{Algo.} & \textbf{Implementation name} & \textbf{Description} \\
        \hline
        1 & \texttt{REoption\_trinomial} & Trinomial tree with ad hoc node initialization followed by standard backward induction \\
        \hline
        2 & \texttt{REoption\_modif\_binomial} & Recursive binomial tree with modified backward induction using equation~\eqref{eq:trinomial-pricing} \\
        \hline
        3 & \texttt{REoption\_modif\_reco\_binomial} & Recombining binomial tree with modified backward induction using equation~ \eqref{eq:trinomial-pricing} \\
        \hline
    \end{tabular}
\end{table}

\begin{table}[htbp]
    \caption{Payoffs used for testing}
    \label{tab:payoff-list}
    \centering
    \begin{tabular}{|c|c|c|c|c|}
        \hline
        \textbf{RE option} & Call & Put & Zero-strike call & Log-contract \\
        \hline
        \textbf{Payoff formula $f(S)$} & $\max(0, S - 100)$ & $\max(0, 100 - S)$ & $S$ &  $\ln(S/100)$ \\
        \hline
    \end{tabular}
\end{table}

\begin{table}[htbp]
    \centering
    \caption{Default pricing parameters}
    \begin{tabular}{|c|c|c|c|c|c|c|c|}
        \hline
        Tree steps $N$ & Maturity $T$ & Spot $ S_0  $ & Volatility $\sigma$ & Interest rate $r$ & Dividend yield $y$ & Expiry intensity $\lambda$ \\
        \hline
        20 & 1 & 100 & 30\% & 10\% & 5\% & 10\%  \\
        \hline
    \end{tabular}
    \label{tab:numerical-params}
\end{table}

\begin{figure}[h]
    \centering
    \includegraphics[width=1\textwidth]{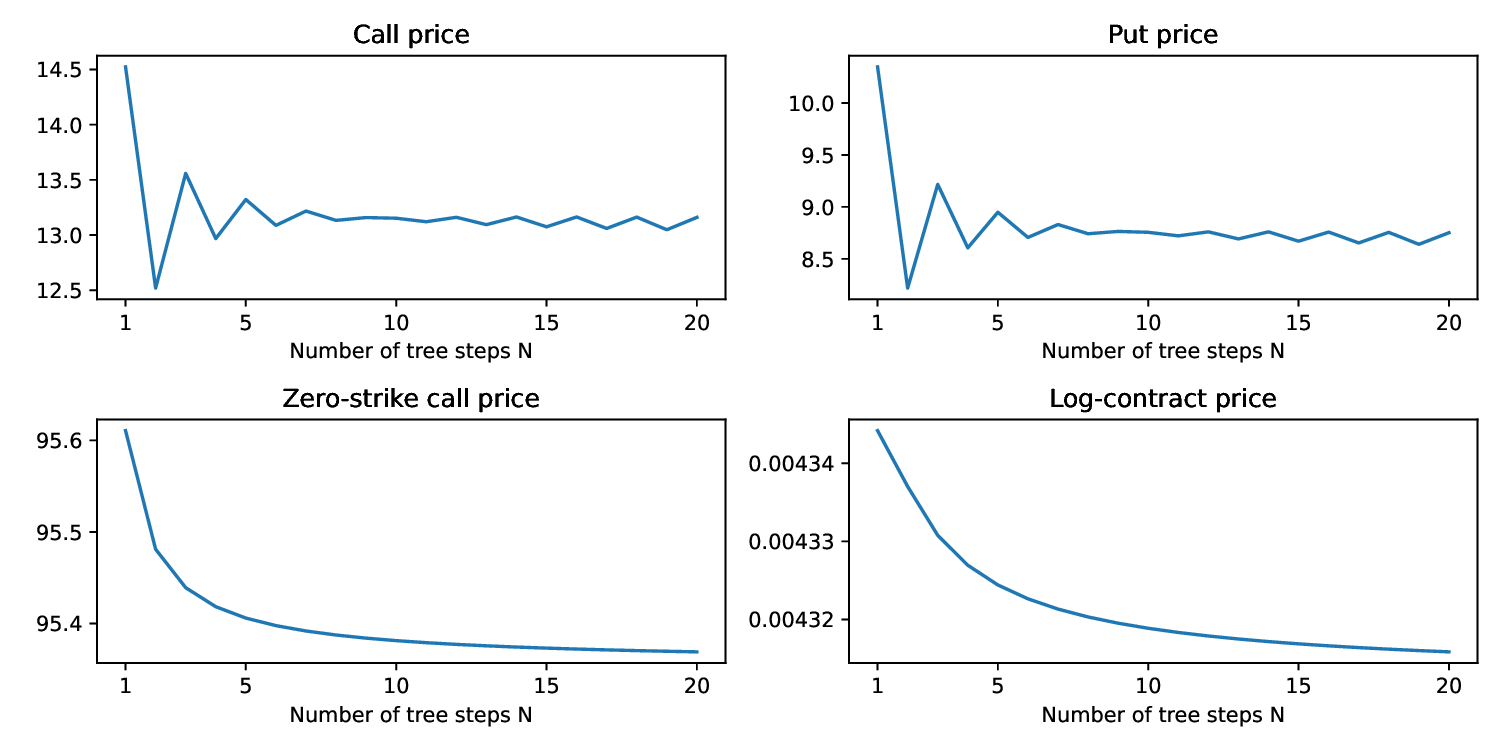}
    \caption{Option price as a function of the number of time periods $N$.}
    \label{fig:opt-price-against-N}
\end{figure}

\begin{figure}[h]
    \centering
    \includegraphics[width=1\textwidth]{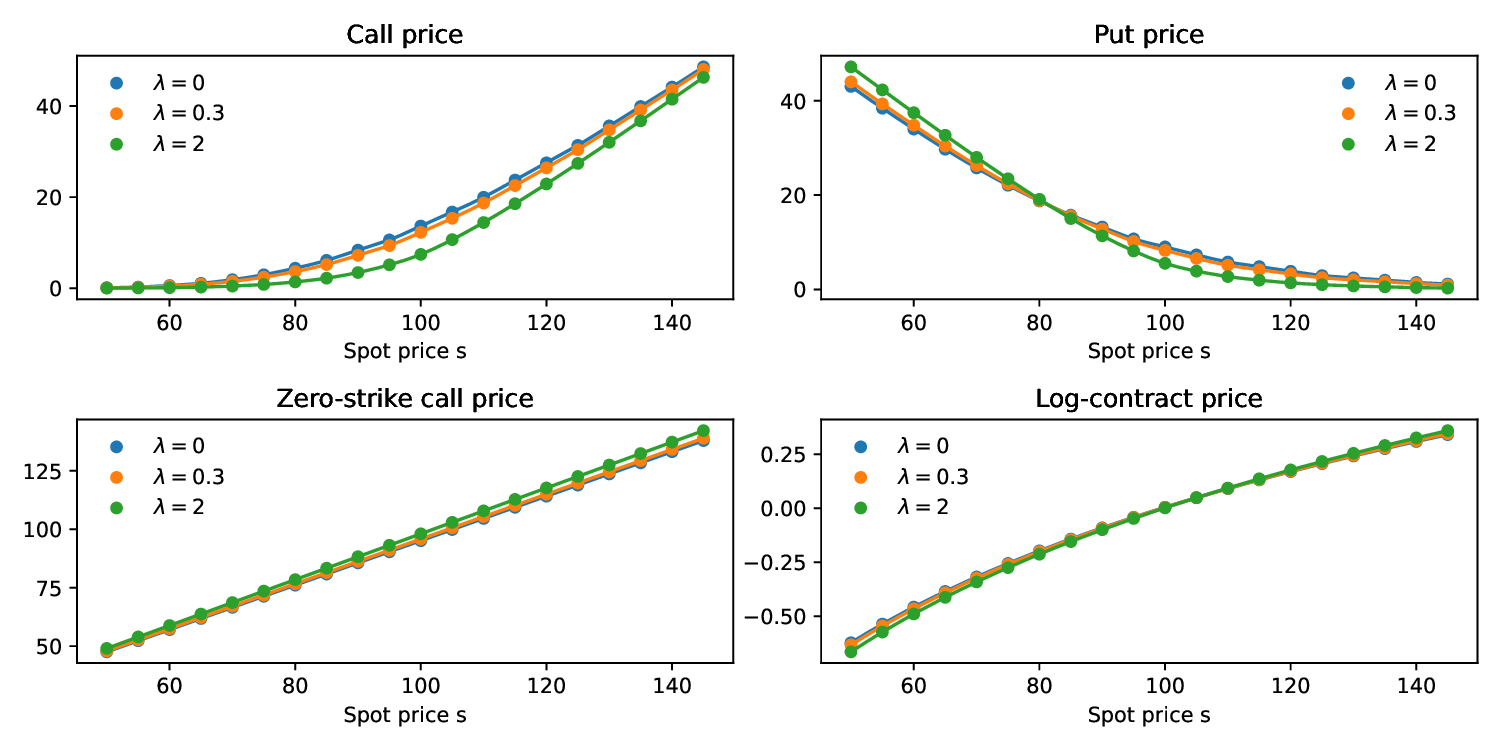}
    \caption{Option price as a function of underlying asset spot price $ S_0  $.}
    \label{fig:opt-price-against-spot}
\end{figure}

\begin{figure}[h]
    \centering
    \includegraphics[width=1\textwidth]{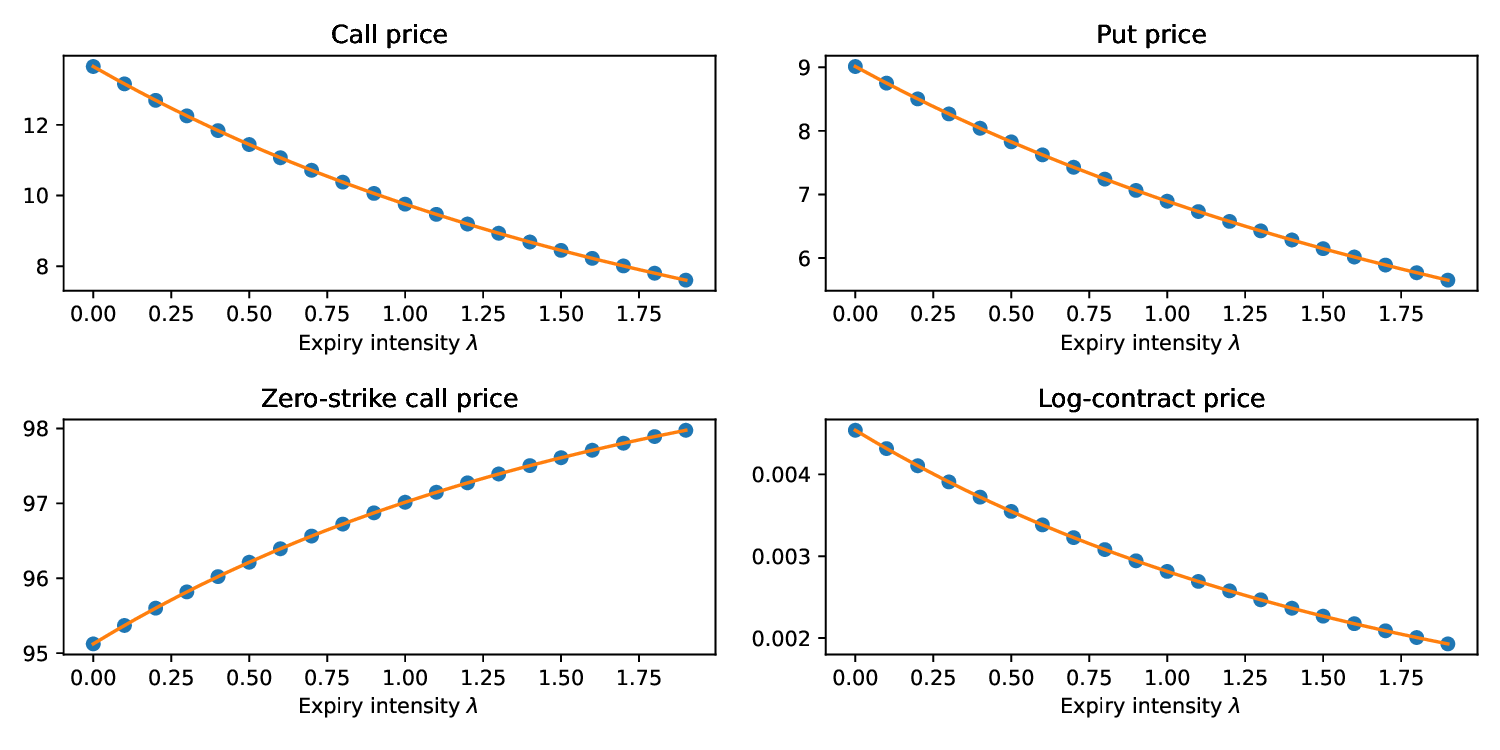}
    \caption{Option price as a function of expiry intensity $\lambda$.}
    \label{fig:opt-price-against-lambda}
\end{figure}

Figures~\ref{fig:opt-price-against-N} through \ref{fig:opt-price-against-lambda} explore the behavior of our RE option pricing methodology.  
Unless otherwise indicated, we used the default parameters listed in Table~\ref{tab:numerical-params}. Figure~\ref{fig:opt-price-against-N} shows the price of each option as a function of the number of time periods $N$.  As expected, all option prices become more stable as $N$ increases, but the call and put prices exhibit oscillations which are typical of tree methods (see, e.g., \cite{leisen-reimer:1996} or \cite{leduc-palmer:2019}).  Figure \ref{fig:opt-price-against-spot} shows the price of each option as a function of the underlying asset spot price $S_0$, for various values of expiry intensity $\lambda$.  As expected, the call and put prices, respectively, increase and decrease with the spot price.  Interestingly, the high-intensity ($\lambda=2$) curve for the put crosses the other two curves around $S_0=80$; this is likely due to the fairly large positive risk-neutral drift $r-y = 5\%$ which makes early termination more favorable when the put is deep in-the-money.  Finally, Figure \ref{fig:opt-price-against-lambda} shows the price of each option as a function of expiry intensity $\lambda$.  We can see that the at-the-money call and put prices decrease as $\lambda$ increases, which is consistent with the intuition that a high chance of early termination reduces the option time value.  In contrast, the RE-ZSC price increases with $\lambda$, which is consistent with \eqref{eq:stock-random-expiry-price}.

\section{Conclusions}   \label{sec:conclusions}

In this paper we introduced RE options. Due to the uncertainty caused by random expiry, these options require novel risk-neutral pricing methods and algorithms.  We showed that a trinomial tree approach where the middle path is interpreted as early expiration can be suitably parameterized to produce arbitrage-free RE option prices.  We also derived the continuous-time limit of this type of tree and made a connection with the expected value of certain functions of subordinated Brownian motion. Finally, our numerical results show how RE options can be priced efficiently using tree-based methods.

A known limitation of risk-neutral pricing in incomplete markets is the lack of a replication strategy in most cases. This can be mitigated, to an extent, by finding a portfolio of traded securities that, in some sense, best approximates the non-replicable payoff, see, e.g., \citet[pp.\ 29--36]{cerny}, \citet{bouzianis-hughston:2020}, and the references therein. An interesting avenue for future research may be to identify approximate static hedges for RE payoffs with the help of neural networks \citep{bossu-crepey-nguyen:2025} and exact static hedges using Radon transforms (\citet{bossu-carr-papanicolaou:2022} and \citet{bossu:2022}).

Additional directions for future research on RE options include: extending to path-dependent RE options (see Section~\ref{sec:trinomial}), adapting other pricing methods such as partial differential equation and Monte-Carlo, and extending our approach to multi-asset RE options using multinomial trees.

\section*{Acknowledgements}
The research of S.~Bossu is partially supported by funds provided by The University of North Carolina at Charlotte. We thank Joseph Bakita for his involvement in an early version of the numerical section.

\appendix

\section{Proofs for Section~\ref{sec:trinomial}} \label{app:proof-th1}

\begin{proof}[Proof of Theorem \ref{thm:main}]

\textbf{(a)}~Taking carried interest into account, the RE payoff $f(S_\tau)$ paid at random expiry $\tau$ is embedded in a standard trinomial tree with terminal payoff $W_T = f(S_\tau)e^{r(T-\tau\Delta t)}$.  The RE option risk-neutral price is then given by
    \[
        e^{-rT} \E^{\Q}(W_T) = e^{-rT} \E^{\Q}\!\big(f(S_\tau)e^{r(T-\tau\Delta t)}\big).
    \]
From here the result follows by simplfying and a standard conditioning argument.

\noindent\textbf{(b)} 
Let $[[S_k]] := \big\{ (\epsilon_1, \dots, \epsilon_k)\in\{u,m,d\}^k : S_0 \epsilon_1\epsilon_2\cdots\epsilon_k = S_k  \big\}$ be the (random) set of all nodes in period $k$ at which the stock price is $S_k$.  We then have \citep[see][Sec.\ 33, for a definition of a probability measure given a random variable]{billingsley:1995}:
\begin{align*}
    \E^{\Q}(\hat S_{k+1} | \hat S_k)
    & =
    \E^{\Q}(e^{(y-r)(k+1)\Delta t} S_{k+1} | S_k)
    = e^{(y-r)(k+1)\Delta t} \E^{\Q}(S_{k+1} | S_k)
    \\
    &= e^{(y-r)(k+1)\Delta t} \sum_{\varsigma\in[[S_k]]}\E^{\Q}\left(S_{k+1} | S_k, \varSigma_k = \varsigma\right)\Q(\varSigma_k = \varsigma | S_k)
    \\
    & =
    e^{(y-r)(k+1)\Delta t} \sum_{\varsigma\in[[S_k]]} \left[S_k u q_u^{(k,\varsigma)} + S_k m q_m^{(k,\varsigma)} + S_k d q_d^{(k,\varsigma)}\right]\Q(\Sigma_k = \varsigma | S_k)
    \\
    & = S_k e^{(y-r)k\Delta t} \sum_{\varsigma\in[[S_k]]} e^{(y-r)\Delta t}\left[u q_u^{(k,\varsigma)} + m q_m^{(k,\varsigma)} + d q_d^{(k,\varsigma)}\right]\Q(\Sigma_k = \varsigma | S_k )
    \\
    & = \hat S_k \sum_{\varsigma\in[[S_k]]}\Q(\Sigma_k = \varsigma | S_k) = \hat S_k,
\end{align*}
where the first equality follows from the definition of $\hat S_{k+1}$ and the fact that $\hat S_{k}$ and $S_k$ generate the same $\sigma$-field, and the last line follows from the fact that $u q_u^{(k,\varsigma)} + m q_m^{(k,\varsigma)}  + d q_d^{(k,\varsigma)} = e^{(r-y)\Delta t}$, as given in \eqref{eq:trinomial-RN}. The martingale property is thus verified. 
  
\noindent\textbf{(c)}~
Consider $1 \leq k \leq  l  < K \leq N-1$ and $x>0$ (the case $l = K$ can be proved similarly). 
By construction,
\begin{align*}
     & \Q(S_k = x , \tau = K\ |\ \tau \geq  l) \\
     =\ & \Q\big(S_k = x, S_{K+1} = m S_K, S_K \neq m S_{K-1}, \dots, S_1 \neq S_0 m \ \big|\ S_l \neq m S_{l-1}, \dots, S_{1} \neq S_0 m \big) \\
     =\ & \frac{ \Q\big( \big[ S_k  = x,  \tfrac{S_l}{S_{l-1}} \neq m, \dots, \tfrac{S_1}{S_0} \neq m  \big] \cap \big[ \tfrac{S_{K+1}}{S_K} = m, \tfrac{S_K}{S_{K-1}} \neq m , \dots, \tfrac{S_{l+1} }{S_l} \neq m \big] \big) }{ \Q(\tau \geq  l ) }.
\end{align*} 
Since the random variables $(\tfrac{S_\kappa}{S_{\kappa-1}})_{1\leq\kappa\leq N}$ are independent under $\Q$ and $S_k = \tfrac{S_k}{S_{k-1}}\cdots \tfrac{S_{1}}{S_0} S_0$, the above expression may be rewritten as
\begin{align*}
     & \Q(S_k = x , \tau = K\ |\ \tau \geq  l) \\
     =\ & \frac{ \Q\big( S_k  = x,  \tfrac{S_l}{S_{l-1}} \neq m, \dots, \tfrac{S_1}{S_0} \neq m  \big) }{ \Q(\tau \geq  l ) }
     \Q\big(\tfrac{S_{K+1}}{S_K} = m, \tfrac{S_K}{S_{K-1}} \neq m , \dots, \tfrac{S_{l+1} }{S_l} \neq m \big)
     \\
     =\ & \Q( S_k  = x\ |\ \tau \geq  l) \Q\big(\tfrac{S_{K+1}}{S_K} = m, \tfrac{S_K}{S_{K-1}} \neq m , \dots, \tfrac{S_{l+1} }{S_l} \neq m \ \big|\ \tau \geq  l  \big),
\end{align*} 
where we used the definition of conditional probability in the last step to obtain the first factor, and independence from $[\tau \geq  l ]$ to obtain the second factor.  It follows that $ \Q(S_k = x , \tau = K \ |\ \tau \geq  l ) = \Q(S_k = x  \ |\ \tau \geq  l  )\Q( \tau = K   \ |\ \tau \geq  l )$, which establishes the required conditional independence. 

\noindent\textbf{(d)}~Note that for any $1\le k\le l$, if we condition on the event $[\tau=l]$, then, with probability $1$, the only possible values of $S_k/S_{k-1}$ are $u$ and $d$. Fix $\epsilon_1,\epsilon_2,\dots,\epsilon_l\in\{u,d\}$ and note that
\begin{eqnarray*}
\Q\left(\frac{S_1}{S_{0}}=\epsilon_1,\dots,\frac{S_l}{S_{l-1}}=\epsilon_l\ \Big|\ \tau=l\right) &=& \frac{\Q\left(\frac{S_1}{S_{0}}=\epsilon_1, \dots,\frac{S_{l}}{S_{l-1}}=\epsilon_l, \frac{S_{l+1}}{S_{l}}= m\right)}{\Q\left(\frac{S_1}{S_{0}}\ne m, \dots,\frac{S_{l}}{S_{l-1}}\ne m, \frac{S_{l+1}}{S_{l}}= m\right)}\\
&=& \prod_{k=1}^l \frac{\Q\left(\frac{S_k}{S_{k-1}}=\epsilon_k\right)}{\Q\left(\frac{S_k}{S_{k-1}}\ne m\right)} \\
&=& \prod_{k=1}^l \frac{\Q\left(e^{X_k}=\epsilon_k\right)(1-q_m^{(k)})}{(1-q_m^{(k)})}\\
&=& \Q\left(e^{X_1}=\epsilon_1,\dots,e^{X_l}=\epsilon_l\right),
\end{eqnarray*}
where the second equality follows by the independence of $\frac{S_1}{S_{0}},\dots, \frac{S_{l+1}}{S_{l}}$, the third by \eqref{eq:RN-homogeneous-trinomial-Nstep-drift} and the definition of $X_k$, and the fourth by the independence of $X_1,X_2,\dots,X_l$. Thus, the conditional distribution of $\left(\frac{S_1}{S_{0}},\dots,\frac{S_l}{S_{l-1}}\right)$ given $[\tau=l]$ is the same as the distribution of $\left(e^{X_1},\dots,e^{X_l}\right)$. 

Since $S_0$ is non-random and $S_k = S_0 \frac{S_1}{S_0} \cdots \frac{S_k}{S_{k-1}}$, it follows that the conditional distribution of $(S_0, S_1, \dots, S_l)$ given $[\tau=l]$ is the same as the distribution of $\left(S_0, S_0 e^{X_1}, \dots, S_0 e^{\sum_{v=1}^l X_v}\right)$. Now, fix $x_1,x_2,\dots,x_N\in\mathbb R$ and $l\in\{0,1,\dots,N\}$. We have
\begin{eqnarray*}
    \Q\left(\tau=l, S_{1\wedge 1}=x_1, \dots, S_{N\wedge l}=x_N \right) &=& \Q\left( S_{1\wedge 1}=x_1, \dots, S_{N\wedge l}=x_N\ |\  \tau = l \right)\Q\left(\tau = l \right)\\
  &=&  \Q\left(  S_0 e^{\sum_{v=1}^{1\wedge l} X_v} = x_1,\dots, S_0 e^{\sum_{v=1}^{N\wedge l} X_v} = x_N \right)\Q\left(\tau = l \right)\\&=& \Q\left(\tau = l, S_0 e^{\sum_{v=1}^{1\wedge \tau} X_v} = x_1,\dots, S_0 e^{\sum_{v=1}^{N\wedge \tau} X_v} = x_N \right),
\end{eqnarray*}
where the last line follows by the fact that the $X_v$'s are independent of $\tau$. Since the joint distributions have the same probability mass functions, the result follows.
\end{proof}

\begin{proof}[Proof of Proposition~\ref{prop:stock-random-expiry}]
Theorem \ref{thm:main}(a) with $f(S)=S$  gives 
    \[
        V_0^{\text{RE-ZSC}} = \E^{\Q}(e^{-r\tau\Delta t} S_\tau) = \sum_{k=0}^N e^{-rk\Delta t} \E^{\Q}(S_k\ |\ \tau =k)  \Q(\tau = k).
    \]
Theorem \ref{thm:main}(d) implies that 
\begin{eqnarray*}
 \E^{\Q}(S_k\ |\ \tau =k) =  \E^{\Q}(S_0 e^{\sum_{ l =1}^{k} X_{ l }}) = S_0\prod_{l=1}^k \E^{\Q}( e^{X_{ l }}) = S_0 e^{(r-y)k\Delta t},
\end{eqnarray*}
where we used the independence of the $X_l$'s and the fact that 
$$
\E^{\Q}(e^{X_{ l }})= u\Q(e^{X_l} =u)+d\Q(e^{X_l}=d) = e^{(r-y)\Delta t}.
$$
It follows that
\[
    V_0^{\text{RE-ZSC}}
    =  \sum_{k=0}^N S_0 e^{-yk\Delta t} \Q(\tau = k) = S_0 \E^{\Q}(e^{-y\tau\Delta t}),
\]
which completes the proof.
\end{proof}

\begin{proof}[Proof of Proposition~\ref{cor:RE-price-range}]
    The first result follows from minimizing and maximizing the expression for $V_0$ in Theorem \ref{thm:main}(a) with respect to the set of all possible distributions for $\tau$. In this case, the maximum and minimum are attained when we take the distribution to be concentrated on one appropriately chosen point. However, such distributions violate the required condition that $\Q(\tau = k) \in(0,1)$ for each $k=0,1,\dots,N$. To comply, we can consider a case close to such a distribution, but with an arbitrarily small amount of mass spread to the other points. This is why we get an open interval.
    The second result follows immediately from the first and the fact that $N\Delta t=T$.
\end{proof}

\section{Proofs for Section~\ref{sec:continuous}}\label{app:proof-th2-3}

\begin{proof}[Proof of Lemma \ref{lem:sup-to-zero}]
Assume that $\tau_n/n\conp\tau_\infty$, fix $\epsilon>0$, let $A_{n,\epsilon} := \big[\left|\tau_n/n-\tau_\infty\right|<\epsilon \big]$, and note that $\Q(A_{n,\epsilon})\to1$ as $n\to\infty$. Note further that $\floor{nt}/n\to t$ and assume that $n$ is large enough that $\left|\frac{\floor{nt}}{n} - t\right| <\epsilon$. Fix $\omega\in A_{n,\epsilon}$ and $t\in[0,T]$.  If $t\le \tau_\infty(\omega)$ and $\floor{nt}\le \tau_n(\omega)$, then
$$
\left|\frac{\tau_n(\omega)\wedge\floor{nt}}{n} -\tau_\infty\wedge t\right| = \left|\frac{\floor{nt}}{n} - t\right| <\epsilon.
$$
If $t\ge \tau_\infty(\omega)$ and $\floor{nt}\ge \tau_n(\omega)$, then
$$
\left|\frac{\tau_n(\omega)\wedge\floor{nt}}{n} -\tau_\infty(\omega)\wedge t\right| = \left|\frac{\tau_n(\omega)}{n} -\tau_\infty(\omega)\right| <\epsilon.
$$
If $t\le \tau_\infty(\omega)$ and $\floor{nt}\ge \tau_n(\omega)$, then $\tau_n(\omega)/n\le \floor{nt}/n \le t\le \tau_\infty(\omega)$ and hence
$$
\left|\frac{\tau_n(\omega)\wedge\floor{nt}}{n} -\tau_\infty(\omega)\wedge t\right| = \left|\frac{\tau_n(\omega)}{n} - t\right| \le  \left|\frac{\tau_n(\omega)}{n} - \tau_\infty(\omega)\right| <\epsilon.
$$
If $t\ge \tau_\infty(\omega)$ and $\floor{nt}\le \tau_n(\omega)$, then if $\tau_\infty(\omega)\le \floor{nt}/n$ we have
$$
\left|\frac{\tau_n(\omega)\wedge\floor{nt}}{n} -\tau_\infty(\omega)\wedge t\right| = \left|\frac{\floor{nt}}{n} -\tau_\infty(\omega)\right|\le \left|\frac{\tau_n(\omega)}{n} - \tau_\infty(\omega)\right| <\epsilon
$$
and if $\tau_\infty(\omega)\ge \floor{nt}/n$ we have
$$
\left|\frac{\tau_n(\omega)\wedge\floor{nt}}{n} -\tau_\infty(\omega)\wedge t\right| = \left|\frac{\floor{nt}}{n} -\tau_\infty(\omega)\right|\le  \left|\frac{\floor{nt}}{n} -t\right|<\epsilon.
$$
Hence, for large enough $n$,
\[
    \left[\sup_{t\in[0,T]}\left|\frac{\tau_n\wedge\floor{nt}}{n} -\tau_\infty\wedge t\right| <\epsilon \right]  \supset A_{n,\epsilon}
\]
and the result follows. For the other direction note that, since the support of $\tau_n$ is contained in $[0,Tn]$ and the support of $\tau_\infty$ is contained in $[0,T]$, we have
\begin{eqnarray*}
     \left|\frac{\tau_n}{n} -\tau_\infty\right|=\left|\frac{\tau_n\wedge\floor{nT}}{n} -\tau_\infty\wedge T\right|\le\sup_{t\in[0,T]}\left|\frac{\tau_n\wedge\floor{nt}}{n} -\tau_\infty\wedge t\right|\conp 0
 \end{eqnarray*}
 as required.
\end{proof}

\begin{lemma}\label{lemma: contin in Skorohod}
Let $f:\mathbb R\to\mathbb R$ be a continuous function and fix $r\ge0$. Let $\psi$ be a mapping from $D([0,T])$ such that for $x\in D([0,T])$, $\psi x$ is a function given by 
$$
\psi x(t) = e^{-rt} f(x(t)), \ \ t\in[0,T].
$$
Then the range of $\psi$ is contained in $D([0,T])$ and $\psi$ is continuous in the Skorohod topology.
\end{lemma}

\begin{proof}
The fact that, for any $x\in D([0,T])$, $\psi x$ is an element of $D([0,T])$ (i.e., that it is right continuous with left limits) follows easily from the fact that $f$ is continuous. Now let $x,x_1,x_2,\dots\in D([0,T])$ such that $x_n\tos x$, where we write $\tos$ to denote convergence in the Skorohod topology. We must show that $\psi x_n\tos \psi x$.

By definition, $x_n\tos x$ means that there exists a sequence $(\lambda_n)$ of nondecreasing continuous functions mapping from $[0,T]$ onto itself and satisfying
$$
\lim_{n\to\infty} \sup_{t\in[0,T]}\left| \lambda_n(t) -t\right|=0 \mbox{ and } \lim_{n\to\infty} \sup_{t\in[0,T]}\left| x_n (\lambda_n(t)) -x(t)\right|=0.
$$
We have
\begin{eqnarray*}
   \left|\psi x_n(\lambda_n(t))- \psi x(t)\right| &=& \left| e^{-r\lambda_n(t)}f(x_n(\lambda_n(t))) - e^{-rt}f(x(t))\right|\\
   &\le& \left| e^{-r\lambda_n(t)}f(x_n(\lambda_n(t))) - e^{-r\lambda_n(t)}f(x(t))\right| \\
   &&\qquad + \left| e^{-r\lambda_n(t)}f(x(t)) - e^{-rt}f(x(t))\right|\\
   &\le& \left|f(x_n(\lambda_n(t))) - f(x(t))\right| + \left| e^{-r\lambda_n(t)} - e^{-rt}\right|\left|f(x(t))\right|.
\end{eqnarray*}
Since $x\in D([0,T])$, there exists a $K>1$ such that $\sup_{t\in[0,T]}|x(t)|\le K-1\le K$, see (12.5) in \cite{Billingsley:1999}. Furthermore, since $x_n\circ\lambda_n$ converges to $x$ uniformly, we have $\sup_{t\in[0,T]}|x_n(\lambda_n(t))|\le K$ for large enough $n$. Since $f$ is continuous, it is uniformly continuous on the compact set $[-K,K]$. Since $x_n \circ\lambda_n\to x$ uniformly, it follows that
$$
\lim_{n\to\infty}\sup_{t\in[0,T]}  \left|f(x_n(\lambda_n(t))) - f(x(t))\right| =0.
$$
For the other part, we have
\begin{eqnarray*}
\lim_{n\to\infty}\sup_{t\in[0,T]} \left| e^{-r\lambda_n(t)} - e^{-rt}\right|\left|f(x(t))\right| \le K \lim_{n\to\infty}\sup_{t\in[0,T]} \left|\int_{r\lambda_n(t)}^{rt} e^{-u}\rd u\right| \le  rK \lim_{n\to\infty}\sup_{t\in[0,T]} \left|\lambda_n(t)-t\right|=0.
\end{eqnarray*}
From here the result follows.
\end{proof}

\begin{proof}[Proof of Theorem \ref{thrm: limits}]
\textbf{(a)}~The proof of Part (a) is influenced by the proof of \cite[Th.\ 14.4]{Billingsley:1999}. Throughout, we write $\cond$ to denote convergence in distribution on the space $D([0,T])$. Since we are only interested in convergence in distribution, for simplicity of notation, we take
 \begin{eqnarray*}
         S_n(t) &=& S(0) \exp\left\{\left(r-y\right)\frac{\floor{nt}}{n} + \sum_{k=1}^{\floor{nt}} \log(1+Y_k \sigma/\sqrt n) \right\}, \ \ t\in[0,T]\\
    S_\infty(t) & =& S(0) e^{ \left(r-y - \frac12 \sigma^2 \right) t + \sigma W_t}, \ \ t\in[0,T],
\end{eqnarray*}
and similarly for $S^*_n(t)$ and $S^*_\infty(t)$, even though, in the statement of the theorem, these only hold as equalities in distribution. Note that all relevant functions of $t$, e.g., $S_n$, $S_n^*$, $\tau_n^*$, and $\tau_\infty^*$, belong to $D([0,T])$, that $\tau_n^*$ and $\tau_\infty^*$ are nondecreasing functions taking values in $[0,T]$, and that $S^*_n = S_n\circ \tau^*_n$ and $S^*_\infty = S_\infty\circ \tau_\infty^*$. 
By standard results \citep[see e.g.][Th.\ 3.2.2 which is easily generalized to nonzero drift]{Shreve:2004}, we have $S_n\cond S_\infty$. By Lemma \ref{lem:sup-to-zero},
\begin{eqnarray*}
    \sup_{t\in[0,T]}\left|\tau_n^*(t) -\tau_\infty^*(t)  \right| \conp 0,
\end{eqnarray*}
which implies convergence in probability of the Skorohod distance to $0$, see \citet[p.124]{Billingsley:1999}. By \citet[Corol.~p.28]{Billingsley:1999}, it follows that $\{\tau_n^*(t):t\in[0,T]\} \cond \{\tau^*_\infty(t):t\in[0,T]\}$. By independence, we get the joint convergence $(\tau_n^*, S_n)\cond (\tau_\infty^*, S_\infty)$ \citep[see][Th.~2.8]{Billingsley:1999}.  Since $S^*_\infty$ is continuous with probability $1$, by \citet[Lemma p.151]{Billingsley:1999}, we get
\[
    S_n^* = S_n\circ \tau^*_n \cond S_\infty\circ \tau_\infty^* = S^*_\infty,
\]
as required.  Turning our attention to the second convergence in Part~(a), for $t\in[0,T]$, let $S_n'(t) = e^{-rt}f(S_n(t))$ and $S_\infty'(t) = e^{-rt}f(S_\infty(t))$. Combining the fact that $S_n\cond S_\infty$ with the mapping theorem (Theorem 2.7 in \cite{Billingsley:1999}) and Lemma \ref{lemma: contin in Skorohod} gives $S_n'\cond S_\infty'$. From here, the proof follows as for the first convergence.

\noindent\textbf{(b)}~Combining Part~(a) with Skorohod's representation theorem implies that there exists a probability space $(\Omega',\mathcal F',\Q')$ and random variables on this space $A_1,A_2,\dots,A_\infty$ satisfying $A_\infty\eqd e^{-r\tau_\infty} f(S_\infty(\tau_\infty))$, $A_n\eqd e^{-r\tau_n/n} f(S_n(\tau_n)$ for each $n=1,2,\dots$, and $A_n\to A_\infty$ with $\Q'$-probability $1$. The fact that $f$ is bounded implies that there is an $M>0$ such that, with probability $1$, $|A_n|\le M$ for each $n=1,2,\dots$. From here the result follows by Lebesgue's dominated convergence theorem.
\end{proof}

\section{Python code implementation}    \label{app:code}

\noindent For ease of reading, non-essential code is omitted and replaced by a jagged line \zigzag{1.8em}. The full code can be found on GitHub.

\lstdefinestyle{mystyle}{
    backgroundcolor=\color{backcolor},   
    commentstyle=\color{codegreen},
    keywordstyle=\color{blue}, 
    numberstyle=\tiny\color{codegrey},
    stringstyle=\color{codepurple},
    basicstyle=\ttfamily\footnotesize,
    escapeinside={(*@}{@*)},
    breakatwhitespace=false,         
    breaklines=true,                 
    captionpos=b,                    
    keepspaces=true,                 
    numbers=left,                    
    numbersep=5pt,                  
    showspaces=false,                
    showstringspaces=false,
    showtabs=false,                  
    tabsize=2,
    emph={None},
    emphstyle={\color{magenta}}
}

\begin{lstlisting}[language=Python,style=mystyle]
#License: CC BY-NC 4.0: Attribution-NonCommercial 4.0 Int'l ((*@\href{https://creativecommons.org/licenses/by-nc/4.0/}{creativecommons.org/licenses/by-nc/4.0}@*))
#Tree pricing of random-expiry (RE) options
#(c) 2025 S. Bossu and M. Grabchak

import numpy
from numpy import isnan
from math import exp, sqrt, log

f = b = u = m = d = qu = qm = qd = None  # global variables

class GlobalPricingParameters:
  def __init__(self):
    self._N = self._T = self._lambda = self._r = self._y = self._sigma = None 
(*@
\hspace{1.5em}\zigzag{15.8} @*)
  def reset(self, N, T, Lambda, r, y, sigma, f=None):
    if f is not None: self.f = f
    self._N, self._T, self._lambda, self._r, self._y, self._sigma = N, T, Lambda, r, y, sigma
    global b, u, m, d, qu, qm, qd
    dt = T/N
    u=exp((r-y)*dt+sigma*sqrt(dt))
    d=exp((r-y)*dt-sigma*sqrt(dt))
    m=exp((r-y)*dt)
    b=exp(-r*dt)
    qm = Lambda * dt
    qu = (m-d)/(u-d)*(1-qm)
    qd = (u-m)/(u-d)*(1-qm)
(*@
\hspace{2.5em}\zigzag{15.5} @*)
globalParams = GlobalPricingParameters()

def stock_trinomial_tree(S):    # trinomial tree of stock prices
  N = globalParams.N
  stockTree = numpy.empty((3**(N+1)-1)//2)
  stockTree[0] = S  #root node
  for i in range((3**N-1)//2):
    stockTree[3*i+1] = d * stockTree[i]
    stockTree[3*i+2] = m * stockTree[i]
    stockTree[3*i+3] = u * stockTree[i]
  return stockTree

def REoption_trinomial(stockTree):     # Algo. 1: trinomial tree of RE option prices
  N = globalParams.N
  optionTree = numpy.full((3**(N+1)-1)//2, numpy.nan, dtype=float)
  def _initTree(j,k):  #initialize middle nodes and descendents
    if(k>N): return
    for l in range(3**(N-k)*(j-1) + (3**(N-k+1)-1)//2, 3**(N-k)*(j-3)+(3**(N-k+2)-3)//2+1):
      optionTree[l] = f(stockTree[(j-2)//3]) / b**(N-k+1)
    _initTree(3*j-1,k+1)   #middle child of node above
    _initTree(3*j+5,k+1)   #middle child of node below
  _initTree(2,1)
  for j in range((3**N-1)//2, (3**(N+1)-1)//2):  #initialize terminal up and down nodes
    if isnan(optionTree[j]): optionTree[j] = f(stockTree[j])
  for j in range((3**N-3)//2, -1, -1):     #backward induction loop
    optionTree[j] = b * (qd*optionTree[3*j+1] + qm*optionTree[3*j+2] + qu*optionTree[3*j+3])
  return optionTree

def REoption_modif_binomial(S, M=None): #  Algo. 2: recursive binomial tree with modified backward induction
  if M is None: M = globalParams.N
  elif (M<=0): return f(S)
  return (b*qu*REoption_modif_binomial(S*u, M-1) + qm*f(S) + b*qd*REoption_modif_binomial(S*d, M-1))

def REoption_modif_reco_binomial(S):  #  Algo. 3: recombining binomial tree with modified backward induction
  N = globalParams.N
  optionTree = numpy.empty(N*(N+3)//2 + 1)
  for k in range(N,-1,-1):  #backward induction loop
    for i in range(k+1):
      if(k==N): #terminal node
        optionTree[N*(N+1)//2+i] = f(S * u**i * d**(N-i))
      else:
        optionTree[k*(k+1)//2+i] = b*qu*optionTree[k*(k+1)//2+i+k+2] + qm*f(S * u**i * d**(k-i))  \
                      + b*qd*optionTree[k*(k+1)//2+i+k+1]
  return optionTree
(*@
\zigzag{16} @*)
\end{lstlisting}


\begin{thebibliography}{14}

\bibitem[\protect\citeauthoryear{American Academy of Actuaries}{2022}]{AAA:2022}
American Academy of Actuaries (2022). Insurance-linked securities and catastrophe bonds. Public Policy Issue Paper. Available at \href{https://actuary.org/wp-content/uploads/2024/12/ILS\_20220614.pdf}{actuary.org/wp-content/uploads/2024/12/ILS\_20220614.pdf}

\bibitem[\protect\citeauthoryear{Azzone and Baviera}{2022}]{Azzone:2022}
M.~Azzone and R.~Baviera (2022). Additive normal tempered stable processes for equity derivatives and power-law scaling. {\em Quantitative Finance}, 22(3), 501--518.

\bibitem[\protect\citeauthoryear{Barndorff and Nielsen}{1997}]{Barndorff-Nielsen:1997}
O.E.~Barndorff-Nielsen (1997). Processes of normal inverse Gaussian type. {\em Finance and stochastics}, 2(1), 41--68.

\bibitem[\protect\citeauthoryear{Bertsimas and Tsitsiklis}{1997}]{bertsimas-tsitsiklis:1997}
D.~Bertsimas and J.N.~Tsitsiklis (1997). {\em Introduction to Linear Optimization}.  Athena Scientific.

\bibitem[\protect\citeauthoryear{Billingsley}{1995}]{billingsley:1995}
P.~Billingsley (1995). {\em Probability and Measure}, 3rd ed. John Wiley \& Sons.

\bibitem[\protect\citeauthoryear{Billingsley}{1999}]{Billingsley:1999}
P.~Billingsley (1999). {\em Convergence of Probability Measures}, 2nd ed. John Wiley \& Sons.

\bibitem[\protect\citeauthoryear{Bossu}{2014}]{bossu:2014}
S.~Bossu (2014).  {\em Advanced Equity Derivatives: Volatility \& Correlation}.  John Wiley \& Sons.

\bibitem[\protect\citeauthoryear{Bossu}{2022}]{bossu:2022}
S.~Bossu (2022). Static replication of European multi-asset options with homogeneous payoff. {\em Applied Mathematical Finance}, 28(5), 381--394.

\bibitem[\protect\citeauthoryear{Bossu, Carr and Papanicolaou}{2022}]{bossu-carr-papanicolaou:2022}
S.~Bossu, P.~Carr and A.~Papanicolaou (2022).
Static replication of European standard dispersion options.
{\em Quantitative Finance}, 22(5), 799--811.

\bibitem[\protect\citeauthoryear{Bossu, Crepey and Nguyen}{2025}]{bossu-crepey-nguyen:2025}
S.~Bossu, S.~Cr\'epey, and H.D.~Nguyen (2025). Spanning multi-asset payoffs with ReLUs. {\em Mathematical Finance}, 35(3), 682--707.

\bibitem[\protect\citeauthoryear{Bossu and Henrotte}{2014}]{bossu:2012}
S.~Bossu and Ph.~Henrotte (2012). {\em An Introduction to Equity Derivatives: Theory \& Practice}, 2nd ed. John Wiley \& Sons.

\bibitem[\protect\citeauthoryear{Bouzianis and Hughston}{2020}]{bouzianis-hughston:2020}
G.~Bouzianis and L.P.~Hughston (2020). Optimal hedging in incomplete markets. {\em Applied Mathematical Finance}, 27(4), 265–-287.

\bibitem[\protect\citeauthoryear{Boyle}{1986}]{Boyle:1986}
P.~Boyle (1986). Option valuation using a three-jump process, {\em International Options Journal}, 3, 7--12.

\bibitem[\protect\citeauthoryear{\v Cern\'y}{2009}]{cerny}
A.~\v{C}ern{\'y} (2009). {\em Mathematical Techniques in Finance: Tools for Incomplete Markets}, 2nd ed. Princeton University Press.

\bibitem[\protect\citeauthoryear{Cont and Tankov}{2004}]{Cont:Tankov:2004}
R.~Cont and P.~Tankov (2004). {\em Financial Modelling with Jump Processes}. Chapman and Hall.

\bibitem[\protect\citeauthoryear{Cox, Ross, and Rubinstein}{1979}]{Cox-Ross-Rubinstein:1979}
J.C.~Cox, S.A.~Ross, and M.~Rubinstein (1979). Option pricing: A simplified approach. {\em Journal of Financial Economics}, 229--263.

\bibitem[\protect\citeauthoryear{Duffie}{1996}]{Duffie:1996}
D.~Duffie (1996).  Incomplete security markets with infinitely many states: An introduction. {\em Journal of Mathematical Economics}, 26(1), 1--8.

\bibitem[\protect\citeauthoryear{Eriksson, Ghysels, and Wang}{2009}]{Eriksson:2009}
A.~Eriksson, E.~Ghysels, and F.~Wang (2009). The normal inverse Gaussian distribution and the pricing of derivatives. {\em Journal of Derivatives}, 16(3), 23.

\bibitem[\protect\citeauthoryear{Grabchak}{2016}]{Grabchak:2016}
M.~Grabchak (2016). {\em Tempered Stable Distributions: Stochastic Models for Multiscale Processes}. Springer.

\bibitem[\protect\citeauthoryear{Grabchak and Samorodnitsky}{2010}]{Grabchak:Samorodnitsky:2010}
M.~Grabchak and G.~Samorodnitsky (2010). Do financial returns have finite or infinite variance? A paradox and an explanation. {\em Quantitative Finance}, 10(8):883--893.

\bibitem[\protect\citeauthoryear{Grigorian and Jarrow}{2024}]{grigorian-jarrow:2024}
K.~Grigorian and R.A.~Jarrow (2024).  Option pricing in an incomplete market.  {\em The Quarterly Journal of Finance}, 14(03).

\bibitem[\protect\citeauthoryear{Ishwaran, Jahandideh, and Zarepour}{2008}]{Ishwaran:Jahandideh:Zarepour:2008}
H.~Ishwaran, M.T.~Jahandideh, and M.~Zarepour (2008). Option pricing for infinite variance data. {\em Statistics}, 42(3), 245---260.

\bibitem[\protect\citeauthoryear{Jacod and Protter}{2017}]{Jacod-Protter:2017}
J.~Jacod and P.~Protter (2017). Option prices in incomplete markets. {\em ESAIM: Proceedings and Survey}, 56, 72--87.

\bibitem[\protect\citeauthoryear{Jarrow}{2023}]{Jarrow:2023}
R.A.~Jarrow (2023). The no-arbitrage pricing of non-traded assets. {\em Annals of Finance}, 19, 401--418.

\bibitem[\protect\citeauthoryear{Karandikar and Rachev}{1995}]{Karandikar:Rachev:1995}
R.L.~Karandikar and S.T.~Rachev (1995).
\newblock  A generalized binomial model and option pricing formulae for subordinated stock-price processes.
\newblock {\em Probability and Mathematical Statistics}, 15:427---447.

\bibitem[\protect\citeauthoryear{K\"ellezi and Webber}{2004}]{Kellezi:Webber:2004}
E.~K\"ellezi and N.~Webber (2004). Valuing Bermudan options when asset returns are L\'evy processes. {\em Quantitative Finance}, 4(1):87--100.

\bibitem[\protect\citeauthoryear{Kim et al.}{2019}]{kim-stoyanov-rachev-fabozzi:2019}
Y.S.~Kim, S.~Stoyanov, S.~Rachev, and F.J.~Fabozzi (2019).  Enhancing binomial and trinomial equity option pricing models.  {\em Finance Research Letters}, 28, 185--190.

\bibitem[\protect\citeauthoryear{Leduc and Palmer}{2019}]{leduc-palmer:2019}
G.~Leduc and K.J.~Palmer (2019). Path independence of exotic options and convergence of binomial approximations. {\em Journal of Computational Finance}, 23(2):73--102. 

\bibitem[\protect\citeauthoryear{Leisen and Reimer}{1996}]{leisen-reimer:1996}
D.P.J.~Leisen and M.~Reimer (1996). Binomial models for option
valuation--examining and improving convergence. {\em Applied Mathematical Finance}, 3(4):319--346.

\bibitem[\protect\citeauthoryear{Longstaff}{1990}]{longstaff:1990}
F.A.~Longstaff (1990). Pricing options with extendible maturities: Analysis and applications. {\em The Journal of Finance}, 45(3), 935--957.

\bibitem[\protect\citeauthoryear{Madan, Carr, and Chang}{1998}]{Madan:Carr:Chang:1998}
D.B.~Madan, P.P.~Carr, and E.C.~Chang (1998). The variance gamma process and option pricing. {\em Review of Finance}, 2(1):79--105.

\bibitem[\protect\citeauthoryear{Madan, Milne, and Shefrin}{1989}]{Madan:Milne:Shefrin:1989}
D.B.~Madan, F.~Milne, and H.~Shefrin (1989). The multinomial option pricing model and its Brownian and Poisson limits. {\em The Review of Financial Studies}, 2(2):251--265.

\bibitem[\protect\citeauthoryear{Maller, Solomon, and Szimayer}{2006}]{Maller:Solomon:Szimayer:2006}
R.A.~Maller, D.H.~Solomon, and A.~Szimayer (2006). A multinomial approximation for American option prices in L\'evy process models. {\em Mathematical Finance}, 16(4):613--633.

\bibitem[\protect\citeauthoryear{McIntosh and Harmetz}{2002}]{mcintosh-harmetz:2002}
C.M.~McIntosh and L.S~Harmetz (2002). Accelerated Vesting of Employee Stock Options: Principles and Strategies. {\em Compensation \& Benefits Review}, 34(2):60--64.

\bibitem[\protect\citeauthoryear{Rachev and Mittnik}{2000}]{Rachev:Mittnik:2000}
S.~Rachev and S.~Mittnik (2000).
\newblock {\em Stable Paretian Models in Finance}.
\newblock Wiley, New York.

\bibitem[\protect\citeauthoryear{Rachev and Ruschendorf}{1995}]{Rachev:Ruschendorf:1995}
S.~Rachev and L.~Ruschendorf (1995).
\newblock Models for option prices.
\newblock {\em Theory of Probability and Its Applications}, 39(1):120--152.

\bibitem[\protect\citeauthoryear{Rachev and Samorodnitsky}{1993}]{Rachev:Samorodnitsky:1993}
S.~Rachev and G.~Samorodnitsky (1993). Option pricing formulae for speculative prices modelled by subordinated stochastic processes. 
\newblock {\em Serdica}, 19:175-190.

\bibitem[\protect\citeauthoryear{Rejman, Weron, and Weron}{1997}]{Rejman:Weron:Weron:1997}
A.~Rejman, A.~Weron and R.~Weron (1997).
\newblock Option pricing proposals under the generalized hyperbolic model.
\newblock {\em Communications in Statistics -- Stochastic Models }, 13(4):867--885.

\bibitem[\protect\citeauthoryear{Sabino}{2023}]{Sabino:2023}
P.~Sabino (2023). Normal tempered stable processes and the pricing of energy derivatives. {\em SIAM Journal on Financial Mathematics}, 14(1), 99--126.

\bibitem[\protect\citeauthoryear{Shreve}{2004}]{Shreve:2004}
S.E.~Shreve (2004). {\em Stochastic Calculus for Finance II: Continuous-Time Models}. Springer.

\bibitem[\protect\citeauthoryear{Wolczynska}{1998}]{Wolczynska:1998}
G.~Wolczynska (1998). Option pricing in incomplete discrete markets. {\em Applied Mathematical Finance}, 5(3---4), 165---179.

\end{thebibliography}
\end{document}